\documentclass[notitlepage, 12pt]{article}

\usepackage{amssymb,amsmath,amsfonts,eurosym,geometry,ulem,graphicx,caption,color,setspace,sectsty,comment,footmisc,caption,natbib,pdflscape,array,hyperref,subcaption,threeparttable,threeparttablex, amsthm, rotating,chngcntr,pdflscape,subfloat, pifont, mathtools, csquotes, longtable}

\newcolumntype{L}{>{\centering\arraybackslash}m{1cm}}
\newcolumntype{C}[1]{>{\centering\let\newline\\arraybackslash\hspace{0pt}}m{#1}}
\newcolumntype{R}[1]{>{\raggedleft\let\newline\\arraybackslash\hspace{0pt}}m{#1}}

\begin{document}

\begin{titlepage}
\title{Social Media Emotions and IPO Returns}

\author{Domonkos F. Vamossy\thanks{email: vamossyd@gmail.com.}
}
\date{\today}
\maketitle

\noindent

\begin{abstract}
	\singlespacing
I examine potential mechanisms behind two stylized facts of initial public offerings (IPOs) returns. By analyzing investor emotions expressed on StockTwits and Twitter, I find that emotions conveyed through these social media platforms can help explain the mispricing of IPO stocks. The abundance of information and opinions shared on social media can generate hype around certain stocks, leading to investors' irrational buying and selling decisions. This can result in an overvaluation of the stock in the short term but often leads to a correction in the long term as the stock's performance fails to meet the inflated expectations. In particular, I find that IPOs with high levels of pre-IPO enthusiasm tend to have a significantly higher first-day return of 29.73\%, compared to IPOs with lower levels of pre-IPO investor enthusiasm, which have an average first-day return of 17.59\%. However, this initial enthusiasm may be misplaced, as IPOs with high pre-IPO investor enthusiasm demonstrate a much lower average long-run industry-adjusted return of -8.22\%, compared to IPOs with lower pre-IPO investor enthusiasm, which have an average long-run industry-adjusted return of -0.14\%. Diving deeper into the qualitative aspects of investor discourse, I find that messages rich in financial language or that bolster prevailing information drive my results. Additionally, a trend towards caution emerges among users who frequently engage with IPOs, perhaps a byproduct of lessons from past endeavors. Intriguingly, firms that enjoy high levels of pre-IPO optimism consistently garner post-launch enthusiasm, a trend at odds with their long-term under-performance.


\vspace{0in}
 \bigskip
\end{abstract}

\setcounter{page}{0}
\thispagestyle{empty}
\end{titlepage}
\pagebreak \newpage

\doublespacing

\clearpage

\section{Introduction} \label{sec:introduction}

Emotions play a crucial role in financial markets, as illustrated by Warren Buffett's adage, ``be fearful when others are greedy, and greedy when others are fearful." This connection is particularly evident in the case of initial public offerings (IPOs), where hype and market sentiment can cause over- or undervaluation of stocks. In particular, the effects of these misvaluations can be seen in the established patterns of IPO returns, such as the typical high first-day returns observed in IPOs (\cite{loughran2002don}) and the inconsistent long-term performance, where small-growth IPOs tend to underperform compared to non-IPO seasoned companies (\cite{loughran1995new}).

The meteoric rise of social media has not only transformed the way we communicate but also how we make investment decisions. Platforms like Twitter and StockTwits have emerged as influential forums where investors, both professional and novice, exchange views, strategies, and predictions about stocks. With millions of messages disseminated daily, these platforms offer rich insights on investor emotions, capturing real-time shifts in market perception. This study seeks to unpack the relationship between such emotions and IPO dynamics.

Building upon previous research on investigating the underlying mechanisms driving IPO returns, this paper introduces a novel approach by directly analyzing emotions expressed by investors on social media platforms. Using investor emotions instead of indirect proxies, such as pre-IPO valuation, provides a more accurate and comprehensive understanding of investor emotions towards a particular stock. In particular, the use of investor emotions as a measure can capture the full range of emotions, provide real-time information, identify patterns and trends in emotions, and provide deeper insight into the underlying mechanisms driving IPO returns. The objective of the research is to investigate whether emotions expressed by investors on social media can help explain first-day IPO returns and subsequent under-performance, adding to the existing literature on IPO returns. Additionally, I explore the dynamics of investor emotions after the IPO, and the persistence of investor emotions over time. 

In this paper, I use a dataset comprising of 187,662 messages from Twitter and 42,757 messages from StockTwits, which is a social media platform for investors to exchange opinions about stocks. The messages in the dataset span from January 2010 to September 2021. The data contains firm-specific messages, which allows me to compute firm-specific emotions. To ensure robust and accurate results, I utilize EmTract, a state-of-the-art open-source python package, which uses natural language processing techniques to identify seven emotional states in the investor messages: neutral, happy, sad, anger, disgust, surprise, and fear.\footnote{The emotions in this paper correspond to the seven emotional states specified in \cite{breaban2018emotional}. For a detailed description of the classification schemes, see \cite{emtract}.} These emotions provide me with a more detailed and accurate understanding of investor sentiment, enabling me to identify patterns and trends that may be missed when using a one-dimensional sentiment measure. As an illustration, when I supply the text "not feeling it :)" to the model, it returns a tuple of (0.064, 0.305, 0.431, 0.048, 0.03, 0.038, 0.084), corresponding to (neutral, happy, sad, anger, disgust, surprise, fear) and when I swap the emoticon from :) to :(, and hence the input to "not feeling it :(", the model output changes to (0.05, 0.015, 0.843, 0.009, 0.015, 0.006, 0.062). This illustrates how the model incorporates emoticons in extracting emotions from the text.\footnote{For other applications of deep learning in economics see \cite{albanesi2019predicting} and \cite{vamossyemotions}. For reviews of machine learning applications in economics, see \cite{mullainathan2017machine} and \cite{athey2019machine}.} 

I use EmTract to generate emotion variables by quantifying the content of each message and then average the results across all messages from ninety days before the IPO up till the market opening on the day of the IPO. I also distinguish between different types of messages by using two classification schemes; the first isolates messages conveying information related to earnings, firm fundamentals, or stock trading from general chat, and the second separates messages conveying original information from those disseminating existing information. I then use a predictive regression to test whether emotions predict first-day returns. I take several steps to mitigate estimation concerns, such as ruling out reactive emotions by looking at the impact of emotions before the IPO. I also tackle mis-attribution by using an additional emotion model for robustness checks. 

My analysis focuses on the role of investor emotions in explaining two stylized facts about IPO returns. I document two main findings. First, I find that IPOs with high levels of pre-IPO investor enthusiasm tend to have a significantly higher first-day return of 29.73\%, compared to IPOs with lower levels of pre-IPO investor enthusiasm, which have an average first-day return of 17.59\%. In a regression setting that controls for IPO characteristics, such as offer amount and 1-month past industry returns, I show that a standard deviation increase in pre-IPO enthusiasm translates into a 14.30\% standard deviation increase in first-day returns. However, this initial enthusiasm may be misplaced, as IPOs with high pre-IPO investor enthusiasm demonstrate a much lower average long-run industry-adjusted return of -8.22\%, compared to IPOs with lower pre-IPO investor enthusiasm, which have an average long-run industry-adjusted return of -0.14\%. Even after controlling for first-day return, IPO characteristics and past industry returns, I find that a standard deviation increase in pre-IPO enthusiasm translates into a 12.74\% standard deviation decrease in 12-month industry-adjusted returns.

Refining my focus on the qualitative nuances of investor communication, I observe that messages rich in financial insights and those that reinforce existing information have a marked influence on stock returns. Further, I explore the continuity of pre-IPO enthusiasm at the individual investor level. The data indicates a consistent bullish sentiment among investors engaging with multiple IPOs. However, for users interacting with a broader spectrum of IPOs, there is a clear temporal change, hinting at a move towards caution, perhaps steered by the trends seen in IPO outcomes. Subsequently, I assess the post-IPO sentiment trajectory at the firm's level, illustrating that companies with initial high enthusiasm consistently garner more optimism. This trend stands out, especially when juxtaposed with the long-term under-performance of these entities. This amplified enthusiasm is mirrored in post-IPO conversations, highlighting that firms with strong pre-IPO buzz are more likely to be the topic of discussion.

My main contribution is to literature studying IPO returns. Some studies have examined the influence of institutional investors on IPO returns. For instance, \cite{JIANG201365} employed over-subscription rates for IPOs in Hong Kong to highlight the role that institutional investors play in determining IPO returns. On the other hand, research by \cite{ljungqvist2006hot} and \cite{ritter2002review} argued that the excitement and enthusiasm of retail investors can lead to overvaluation of an IPO stock and its first-day return, which ultimately results in long-term under-performance as the stock reverts to its fundamental value.\footnote{\cite{da2011search} provide an attention-driven explanation. They argue that for an investor to become overly enthusiastic about an upcoming IPO, they must allocate attention to the equity issuance. Without attention from retail investors, these investors are less likely to impact the first-day return and long-term performance of the IPO. For an extensive literature review on IPO underpricing see \cite{ljungqvist2007ipo}.} \cite{cornelli2006investor} further substantiated this observation using pre-IPO pricing data from European markets. What sets my paper apart is its methodology: while these studies predominantly rely on indirect proxies to gauge investor sentiment, my research adopts a more direct approach. Additionally, a unique facet of my work is the examination into the evolution of sentiment at the investor level.

Closest to my setting, \cite{tsukioka2018investor} uses investor sentiment extracted from Yahoo! Japan Finance message boards on 654 Japanese IPOs from 2001 to 2010, and shows that increased investor attention and optimism lead to higher IPO offer prices and initial returns, offering insight into the typical initial high returns followed by the underperformance seen in IPOs. Yet, the study is limited to pre-IPO sentiment in the Japanese context and employs standard machine learning classification models like SVM. According to \cite{emtract}, these classifiers fall short in emotion detection compared with modern language models. In stark contrast, my research harnesses a state-of-the-art language model adept at discerning a comprehensive emotional spectrum, including happiness, sadness, disgust, anger, fear, and surprise. While past research predominantly concentrated on overarching positive and negative sentiment metrics, my study probes deeper to uncover the potential impact of distinct emotions on IPO underpricing. Moreover, my research underscores the qualitative depth of investor communication, noting that messages rich in financial insights significantly influence stock returns. By expanding the dataset beyond the IPO phase, I trace a dynamic sentiment evolution, revealing a consistent bullish trend pre-IPO, a shift towards caution with broader IPO interactions, and sustained optimism for firms, even when they underperform long-term. 

My study also adds to the literature on behavioral finance by showing that firm-specific investor emotions on social media platforms can help explain IPO returns. Previous research has shown that emotions play a role in generating bubbles in experimental asset markets (e.g., \cite{breaban2018emotional}, \cite{andrade2016bubbling}), traders' moods can lead to price movements at the market level (e.g., \cite{kamstra2003winter}, \cite{hirshleifer2003good}, \cite{bollen2011twitter}, \cite{gilbert2010widespread}). However, I use direct proxies by measuring emotions in social media posts directly related to a stock listing. My findings are consistent with laboratory experiments that have shown positive emotions lead to higher prices and larger bubbles.

Another contribution of this paper is to the literature studying social media's role in capital markets. Previous studies have explored how companies use social media as a means of investor communication (e.g., \cite{blankespoor2014role}, \cite{jung2018firms}), and have investigated the predictive power of social media content (e.g., \cite{antweiler2004all}, \cite{das2007yahoo}, \cite{chen2014wisdom}, \cite{mao2012correlating}, \cite{curtis2014investor}, \cite{cookson2020don}). My study builds upon this by showing that emotions extracted from social media can help predict a firm's first-day IPO returns and subsequent long-run under-performance.

The remainder of the paper is organized as follows. Section \ref{sec:data} describes the data; Section \ref{sec:primary} presents my primary results; Section \ref{sec:sensitivity} presents my secondary findings and conducts a sensitivity analysis. Section \ref{sec:conclusion} concludes.

\section{Data}\label{sec:data}

\subsection{Data Sources}

I gathered data on all IPOs of common stocks completed between January 2010 and September 2021 in the United States directly from NASDAQ's website. This dataset includes details such as the IPO date, ticker, price, and the total offer amount. I then supplemented this with price-related variables sourced from the merged CRSP/COMPUSTAT file. 

Next, I collected social media data for companies filing IPOs. This data spans the period starting 90 days prior to the IPO, and concludes on the IPO day itself, ceasing at 9:29 am. My primary focus was on messages that could be definitively attributed to specific stocks. Hence, I narrowed my attention to those messages that exclusively mentioned one ticker. To further refine the quality of this social media data, I set a threshold: only stocks with a minimum of two posts prior to the IPO were considered, as this helps to filter out noisy signals.

In defining my sample, I implemented the restrictions for IPO firms as proposed by \cite{da2011search}. Accordingly, I included only those common stock IPOs priced at \$5 or above, which were traded on the NYSE, AMEX, or NASDAQ, and for which the first available CRSP closing price was recorded within five days post-IPO. Last, I excluded both IPO and non-IPO firms with a market capitalization below \$25M to mitigate liquidity concerns.

\subsubsection*{StockTwits Data}

I constructed a sample of investor emotions extracted from StockTwits, a popular social networking platform for investors, which allows users to share their stock opinions. The platform is similar to Twitter, where users can post messages of up to 140 characters (or 280 characters since late 2019) and use ``cashtags" with the stock ticker symbol (e.g., \$AMZN) to link their ideas to a specific company. While StockTwits does not directly integrate with other social media platforms, participants can share content to their personal Twitter, LinkedIn, and Facebook accounts.

For the StockTwits data, for each message, I observe sentiment indicators tagged by the user (bullish, bearish, or unclassified), sentiment score computed by StockTwits, ``cashtags'' that connect the message to particular stocks, like count, and a user identifier. The user identifier allows me to explore user characteristics, such as follower count. For most users, I also have information on a self-reported investment philosophy that can vary along two dimensions: (1) Approach - technical, fundamental, momentum, value, growth, and global macro; or (2) Holding Period - day trader, swing trader, position trader, and long term investor. Users of the platform also provide their experience level as either novice, intermediate, or professional. This user-specific information about the style, experience, and investment model employed helps explore heterogeneity in investor emotions. After my sample restrictions, I ended up with 42,757 messages, authored by 16,183 users, covering 745 IPOs.

Through StockTwits, I can analyze the emotions expressed by investors in the period following the IPO event. Paying particular attention to  the evolution of these emotions — how they change, intensify, or diminish—over a year after the IPO. This extended time-frame offers critical insights into the sustained sentiment of investors as they react to the company's post-IPO trajectory, influenced by its performance, news, and broader market events. What further enriches this study is the detailed data on user profiles from StockTwits, which encompasses information on experience, trading horizon, and approach. This granularity allows for a deeper exploration of investor heterogeneity. Such comprehensive insights from StockTwits make it the primary focus of my analysis. 

\subsubsection*{Twitter Data}

For robustness, I use a rich dataset of emotions extracted from Twitter, a social media platform that allows users to share their thoughts and opinions. Users can post messages, known as "tweets," of up to 280 characters and use the ``@" symbol to mention other users, the ``\#" symbol to tag topics of interest (e.g., \#Bitcoin), and use ``cashtags'' with the stock ticker symbol (e.g., \$AMZN) to link their opinions to stocks. In total, there are 187,662 messages, authored by 65,225 users, corresponding to 1,490 IPOs. While Twitter offers broader coverage, its data is limited to sentiments expressed until the market opens on the IPO day.

\subsection{Descriptive Statistics}

In the primary sections of my research, I employ the emotion ``Happy" as the primary indicator of ``Investor Enthusiasm." This singular emotion offers a snapshot into the positive sentiment and excitement surrounding an IPO. To bolster the robustness of my findings, I integrate the various emotional responses into a consolidated metric known as ``Valence". This is calculated by contrasting the sentiment of happiness against the cumulative impact of sadness, anger, disgust, and fear. The formulation is consistent with the approach presented by \cite{breaban2018emotional}:

\begin{equation}
\text{Valence} = \text{Happy} - \text{Sad} - \text{Anger} - \text{Disgust} - \text{Fear}
\end{equation}

However, it's essential to underscore that, while ``Investor Enthusiasm" and ``Valence" stand central to my analyses, the individual emotions are not sidestepped. For those desiring a more nuanced understanding, the appendix furnishes a detailed exploration of each emotion, offering insights beyond the broad strokes painted by the ``Happy'' and ``Valence'' metrics.

\begin{figure}[htbp] 

\begin{center}
	\subfloat[]{\includegraphics[scale=0.33]{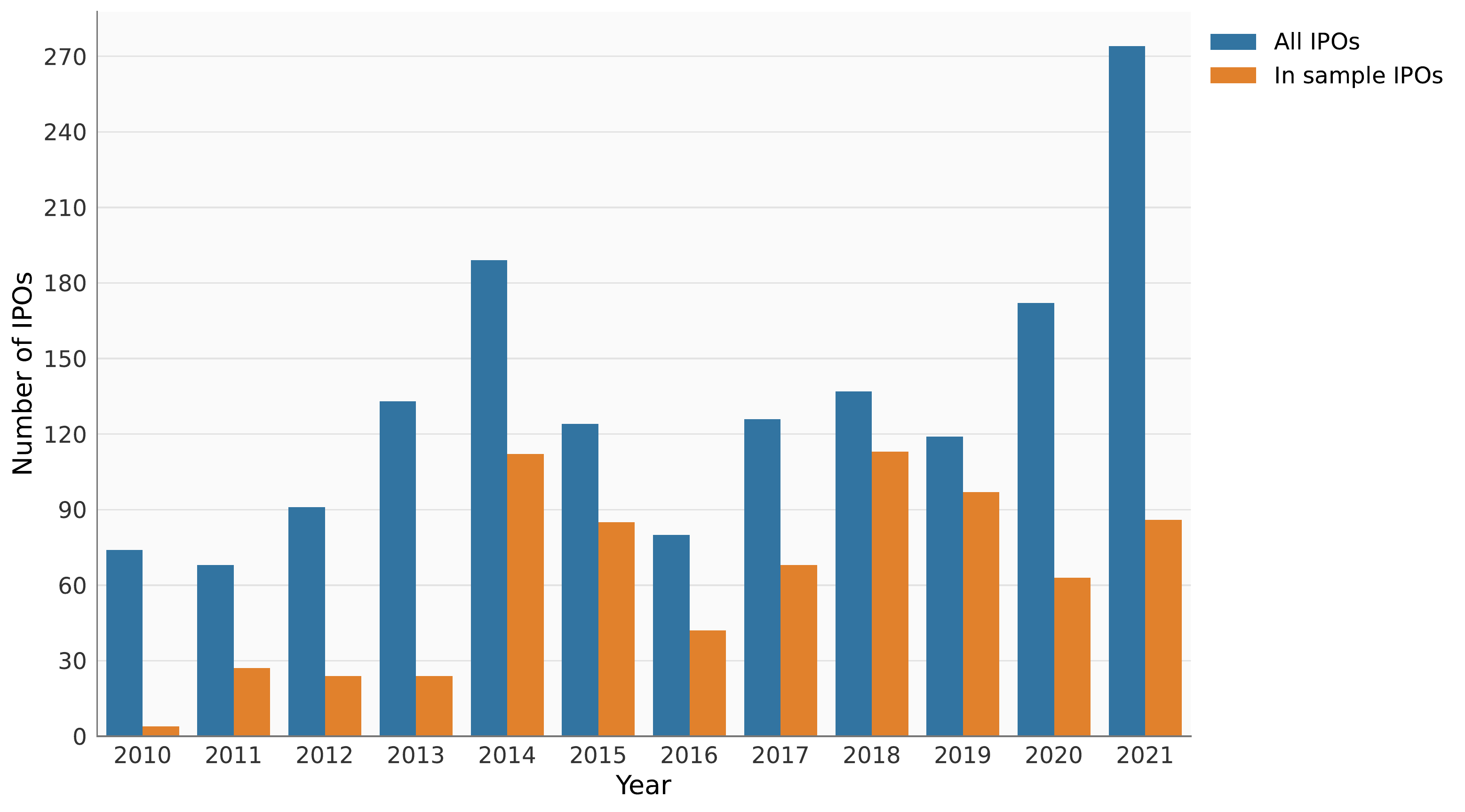}}
 
    \subfloat[]{\includegraphics[scale=0.33]{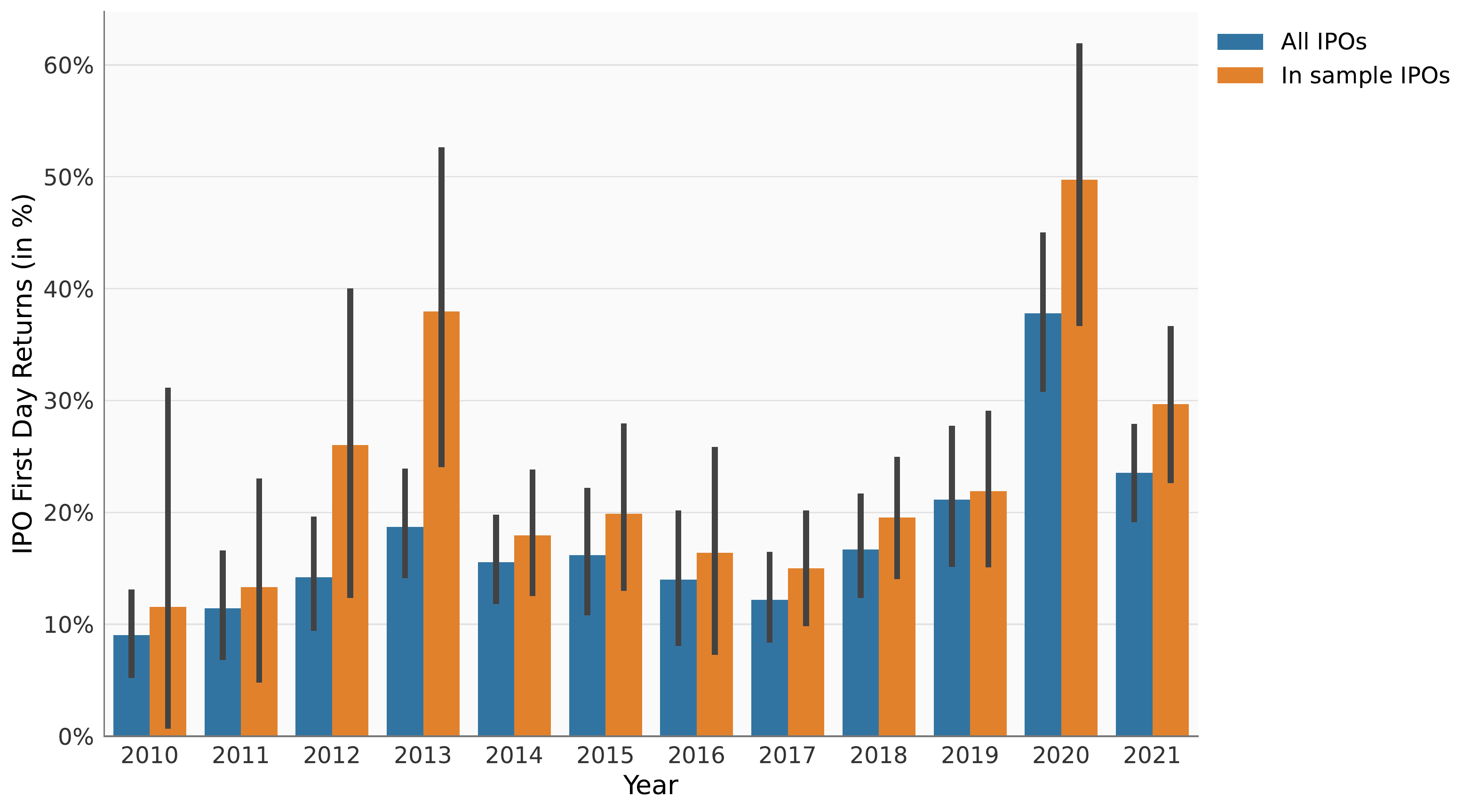}}

    \subfloat[]{\includegraphics[scale=0.33]{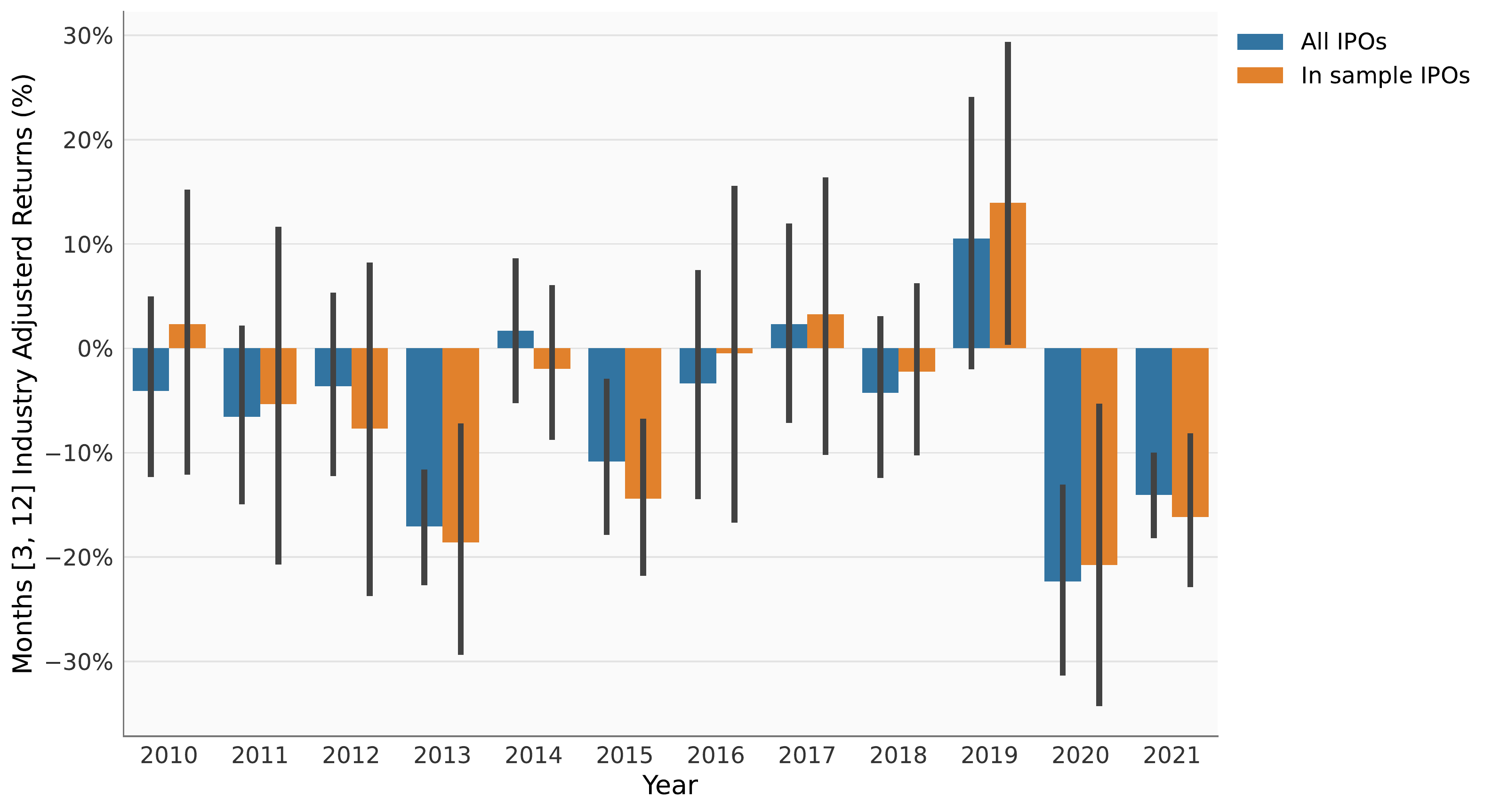}}

	\caption{StockTwits's Coverage of IPOs}\label{fig:ipo_coverage}
\end{center}
 
\begin{flushleft}
	\footnotesize{Notes: StockTwits provides a sample representative of the broader IPO market.}
\end{flushleft}

\end{figure}

Figure \ref{fig:ipo_coverage} explores the coverage of IPOs on StockTwits. In Panel (a), a histogram contrasts the number of IPOs in the matched sample with the overall count of IPOs, demonstrating a significant proportion of IPOs being covered by social media. Notably, there was a dip in this coverage in the study's early years, aligning with the burgeoning growth of social media. Yet, from 2014 to 2019, there was a marked rise in coverage, emphasizing that a predominant share of IPOs finds mention on social media platforms. Panel (b) showcases the average and standard deviation of first-day returns for all IPOs as compared to those featured on StockTwits. Meanwhile, Panel (c) charts the long-run industry-adjusted returns, spanning from 3 months post-IPO to 12 months thereafter. Interestingly, the plots reveal that the average and standard deviation metrics for both IPO categories closely mirror each other, maintaining consistency over time. This alignment strengthens the argument that the sampled IPOs genuinely reflect the broader IPO landscape. The extensive reach of IPOs on social media bolsters the credibility and relevance of my findings for the wider IPO market. For an expanded exploration of StockTwits coverage, refer to Appendix \ref{app:stocktwits_activity}. Additional insights into the coverage from Twitter data can be found in Appendix \ref{app:twitter_activity}.

\begin{table}[htbp]\centering
\begin{threeparttable}
\footnotesize 
\def\sym#1{\ifmmode^{#1}\else\(^{#1}\)\fi}
\caption{Summary Statistics: IPO Sample\label{tab:summary_stats_ipo}}
\begin{tabular}{l*{7}{c}}
\hline\hline
& \multicolumn{7}{c}{CRSP/Compustat-StockTwits-Nasdaq} \\
\hline
& & & & & & & \\[\dimexpr-\normalbaselineskip+2pt] 
 &    Mean &      Std. Dev. &     1\% &    25\% &     50\% &     75\% &      99\% \\ \hline 
& & & & & & & \\[\dimexpr-\normalbaselineskip+2pt] 
\multicolumn{8}{l}{\textbf{Social Media Information}} \\
& & & & & & & \\[\dimexpr-\normalbaselineskip+2pt] 
\textcolor{white}{...} Number of Posts & 57.392 & 245.648 & 2 & 5 & 11 & 29 & 1114 \\
\textcolor{white}{...} Number of Users & 32.034 & 144.554 & 1 & 3 & 6 & 13 & 589.880 \\
\textcolor{white}{...} Happy & 0.160 & 0.140 & 0.008 & 0.036 & 0.133 & 0.246 & 0.607 \\
\textcolor{white}{...} Sad & 0.017 & 0.022 & 0.001 & 0.004 & 0.008 & 0.023 & 0.110 \\
\textcolor{white}{...} Hate & 0.024 & 0.039 & 0.001 & 0.004 & 0.010 & 0.033 & 0.136 \\
\textcolor{white}{...} Surprise & 0.050 & 0.066 & 0.002 & 0.006 & 0.024 & 0.078 & 0.337 \\
\textcolor{white}{...} Fear & 0.048 & 0.053 & 0.006 & 0.016 & 0.033 & 0.068 & 0.215 \\
\textcolor{white}{...} Valence & 0.071 & 0.153 & -0.389 & -0.002 & 0.043 & 0.136 & 0.574 \\
& & & & & & & \\[\dimexpr-\normalbaselineskip+2pt] 
\multicolumn{8}{l}{\textbf{IPO \& Financial Information}} \\
 & & & & & & & \\[\dimexpr-\normalbaselineskip+2pt] 
\textcolor{white}{...} First Day Return & 0.237 & 0.353 & -0.244 & 0.001 & 0.138 & 0.381 & 1.567 \\
\textcolor{white}{...} IPO Price & 18.392 & 13.761 & 5 & 14 & 16.500 & 20 & 54.240 \\
\textcolor{white}{...} IPO Shares & 17.431 & 29.937 & 1.141 & 5.750 & 9.100 & 17.250 & 134.500 \\
\textcolor{white}{...} Offer Amount & 423.031 & 1403.342 & 10.660 & 80 & 146.625 & 331.579 & 3450.400 \\
\textcolor{white}{...} Industry Adjusted Return$_{2m, 12m}$ & -0.064 & 0.507 & -0.948 & -0.381 & -0.140 & 0.132 & 1.825 \\
\textcolor{white}{...} Industry Adjusted Return$_{3m, 12m}$ & -0.048 & 0.500 & -0.928 & -0.366 & -0.124 & 0.149 & 2.003 \\
\textcolor{white}{...} Industry Adjusted Return$_{4m, 12m}$ & -0.062 & 0.443 & -0.862 & -0.352 & -0.123 & 0.135 & 1.610 \\
\textcolor{white}{...} 1-month Lag Industry Return & 0.031 & 0.239 & -0.495 & -0.121 & 0.019 & 0.151 & 0.805 \\ \hline 
& & & & & & & \\[\dimexpr-\normalbaselineskip+2pt] 
\textcolor{white}{...} Observations &    &     & &  &  &    &   754 \\ \hline \hline

\end{tabular}
   \begin{tablenotes}
     \footnotesize
    \item Notes: Offer Amount and IPO Shares in millions. Emotions from 90 days before the IPO date up till 9.29AM of the IPO date; firms with at least two messages. Return$_{2m, 12m}$ corresponds to IPO firm return from two months after the IPO until a year after the IPO. Return variables are winsorized at the 0.5\% and 99.5\% levels. StockTwits data.
   \end{tablenotes}
\end{threeparttable}
\end{table}
 
\begin{sidewaystable}[htbp]\centering
\begin{threeparttable}
\footnotesize
\caption{Correlation Matrix}\label{tab:corr_matrix}
\def\sym#1{\ifmmode^{#1}\else\(^{#1}\)\fi}
\begin{tabular}{l*{9}{c}}
\hline\hline
                &\multicolumn{9}{c}{}                                                                                                                                                      \\
                &  Valence         &    Happy         &      Sad         &     Hate         & Surprise         &     Fear         &Industry Adjusted Return$_{3,12}$        &Offer Amount       &       \\
\hline
& & & & & & & \\[\dimexpr-\normalbaselineskip+2pt] 
Happy           &     0.82\sym{***}&            &                  &                  &                  &                  &                  &                  &                  \\
& & & & & & & \\[\dimexpr-\normalbaselineskip+2pt] 
Sad             &    -0.31\sym{***}&     0.11         &            &                  &                  &                  &                  &                  &                  \\
& & & & & & & \\[\dimexpr-\normalbaselineskip+2pt] 
Hate            &    -0.33\sym{***}&     0.11         &     0.52\sym{***}&            &                  &                  &                  &                  &                  \\
& & & & & & & \\[\dimexpr-\normalbaselineskip+2pt] 
Surprise        &     0.04         &     0.17\sym{***}&     0.17\sym{***}&     0.17\sym{***}&            &                  &                  &                  &                  \\
& & & & & & & \\[\dimexpr-\normalbaselineskip+2pt] 
Fear            &    -0.35\sym{***}&     0.14\sym{**} &     0.40\sym{***}&     0.30\sym{***}&     0.15\sym{**} &            &                  &                  &                  \\
& & & & & & & \\[\dimexpr-\normalbaselineskip+2pt] 
Industry Adjusted Return$_{3,12}$&    -0.07         &    -0.10         &     0.01         &    -0.01         &    -0.06         &    -0.07         &              &                  &                  \\
& & & & & & & \\[\dimexpr-\normalbaselineskip+2pt] 
Offer Amount  &     0.04         &     0.14\sym{**} &     0.12\sym{*}  &     0.16\sym{***}&     0.10         &     0.09         &    -0.02         &              &                  \\
& & & & & & & \\[\dimexpr-\normalbaselineskip+2pt] 
First Day Return&     0.13\sym{*}  &     0.14\sym{**} &    -0.03         &     0.04         &     0.07         &    -0.01         &    -0.09         &     0.13\sym{*}  &       \\
& & & & & & & \\[\dimexpr-\normalbaselineskip+2pt] 
\hline
Observations    &      745         &                  &                  &                  &                  &                  &                  &                  &                  \\ \hline 
\hline\hline
\end{tabular}
\begin{tablenotes}
\scriptsize 
\item Notes: Pairwise correlation with Bonferroni corrections. Return variables are winsorized at the 0.5\% and 99.5\% levels. \textit{t} statistics in parentheses. \sym{*} \(p<0.05\), \sym{**} \(p<0.01\), \sym{***} \(p<0.001\). StockTwits data.
\end{tablenotes}
\end{threeparttable}
\end{sidewaystable}

Table \ref{tab:summary_stats_ipo} presents the descriptive statistics for the analysis variables using the StockTwits sample. The table captures the emotions expressed on StockTwits concerning each IPO, complemented by the relevant IPO and financial data. Conversely, in Appendix \ref{app:social_media_data} Table \ref{tab:summary_stats_ipo_tw} conveys analogous information, drawing from Twitter posts. The panels include the percentage of posts that expressed happiness, sadness, hate, surprise, and fear related to each IPO, as well as the total number of posts and users. I used EmTract to extract emotions from my data and employed two emotion models: one specifically designed for StockTwits, and one general emotion model. 

I found that a large number of messages were categorized as neutral (70.1\% on average), which was partly due to the information shared about the upcoming IPO. Interestingly, Twitter contains a larger fraction of neutral messages (79.1\% for Twitter). 16\% of happy posts follow neutral. Hence, looking at the emotion variables, I observe a positive skewness. This might suggest that investors are more likely to share their enthusiasm on social media than pessimism. Fear and surprise are the third and fourth most frequent emotions, followed by hate and sadness. 

In line with stylized facts about IPO returns, I document high average first-day returns (23.7\% for StockTwits and 20.2\% for Twitter), and negative industry-adjusted returns (i.e., long-run underperformance). Last, although the descriptive statistics are similar between StockTwits and Twitter, the IPO and financial information indicate that StockTwits is more inclined to cover larger IPOs.

Table \ref{tab:corr_matrix} presents pairwise correlation coefficients among my analysis variables for StockTwits, and Table \ref{tab:corr_matrix_tw} presents it for the Twitter sample in Appendix \ref{app:social_media_data}. These include my emotion variables along with first-day return, long-run return, and my main control variable, Offer Amount. My ``happy'' (investor enthusiasm) measure shows statistically significant correlations with my dependent variable, first-day return. This may be viewed as prima facie evidence of the predictive ability of aggregate emotions from individual posts regarding IPO first-day returns. In addition, the relatively small pairwise correlation coefficients among my control variables indicate little evidence of multi-collinearity in my data.\footnote{Furthermore, the high correlation between the emotion variables suggests a large proportion of neutral messages in the dataset.}

\section{Primary Findings}\label{sec:primary}

I examine how changes in emotion affect stock returns and demonstrate that elevated investor enthusiasm, as reflected in social media data, can induce buying pressure, leading to a temporary increase in stock prices. Subsequently, I show that these stocks, after their initial surge driven by this enthusiasm, tend to under-perform when compared to their industry peers. Turning my attention to the qualitative aspects of investor communication, I find that messages rich in financial details and those that amplify existing information have the most pronounced influence on stock dynamics. I also examine the persistence of pre-IPO enthusiasm at the individual investor level. My findings reveal a prevailing bullish sentiment among investors across multiple IPOs. However, this sentiment becomes more varied among users who have engaged with a greater number of IPOs. This varied sentiment is coupled with a noticeable temporal evolution, indicating a shift towards caution, perhaps influenced by patterns observed in IPO performances. Last, I examine the sentiment trajectory post-IPO at the firm level, and show that firms with higher initial enthusiasm consistently attract more optimism. This observation is intriguing, especially considering the long-term under-performance of these firms. This heightened enthusiasm is also reflected in post-IPO discussions, where firms with pronounced pre-IPO enthusiasm are more likely to be discussed.

While I do not establish a causal relationship between social media emotions and asset price movements, my findings highlight the relevance of investor emotions in stock valuation, and provide empirical evidence for two stylized facts about IPO returns.

\subsection{Empirical Framework}

My primary predictive regression is as follows:

\begin{equation}\label{eq:emotion}
Y_{it} = \alpha + \beta_1 \text{Investor Enthusiasm}_{it} + \zeta'X_{it} + \epsilon_{it}
\end{equation}

In this equation, the term Investor Enthusiasm represents my principal emotion measure for firm i at date t, sourced from ``happy'' messages posted within a 90-day window preceding the IPO. My dependent variable alternates between a first-day return and a 12-month industry-adjusted return. One major concern to be addressed is the possibility that the emotion measure merely echoes changes in trading activity. To mitigate this, I have structured the methodology such that the emotion measure always precedes the dependent variable chronologically.

The cyclical nature of IPO markets, as elaborated by \cite{lowry2002ipo}, suggests these patterns may mirror larger economic shifts \cite{pastor2005rational}. Extending this line of thought, \cite{derrien2005ipo} propose that prevailing market conditions can, at times, eclipse intrinsic IPO traits, especially during ``hot" market spells. In light of these insights, my regressions control for the average first-day return of recent IPOs — those within the 90 days preceding firm i's IPO. Additionally, in certain specifications, I incorporate year and industry fixed effects.\footnote{For a granular analysis of emotion measures, see Appendix \ref{app:emotions}. To avoid multi-collinearity, the neutral emotion has been set as the reference category.}

\subsection{Investor Enthusiasm and IPO Pricing}\label{sec:ipo}

I now investigate the underlying mechanism behind two commonly observed patterns in IPO returns. First, IPOs tend to exhibit high first-day returns (as noted in \cite{loughran2004has}). Second, small-growth IPOs often under-perform established non-IPO companies. To directly test \cite{ljungqvist2006hot} conjecture regarding the role of retail investor enthusiasm in driving up first-day returns and causing long-term under-performance, I analyze investor emotions. Specifically, I examine the relationship between first-day IPO returns and investor enthusiasm, measured from ninety days before the IPO date up to 9.29AM of the IPO date. For the second part, I investigate how these emotions interact with first-day IPO returns to better understand their relationship with long-term (under)performance.

\begin{figure}
\begin{center}
    \subfloat[]{\includegraphics[scale=0.45]{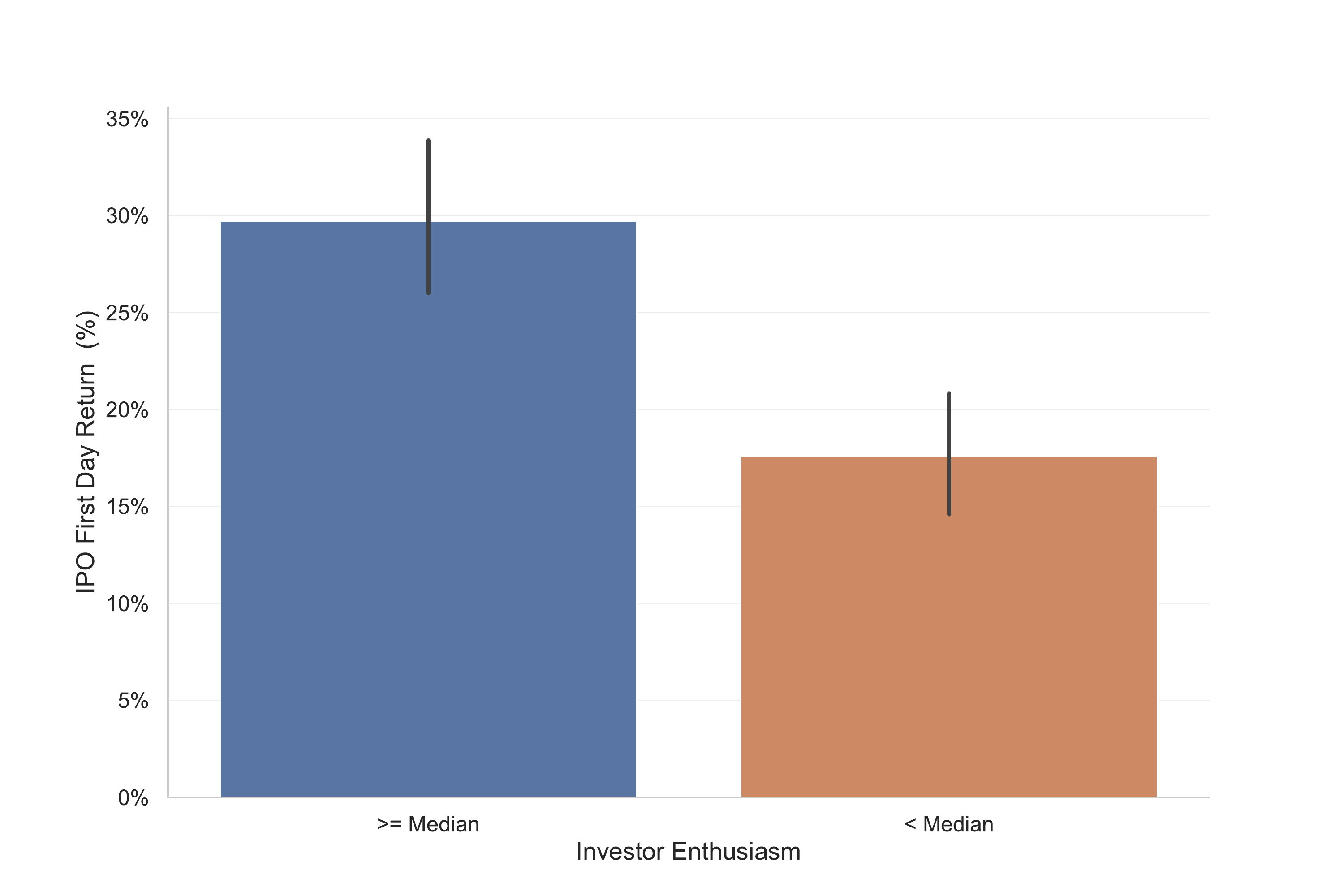}}
    
    \subfloat[]{\includegraphics[scale=0.45]{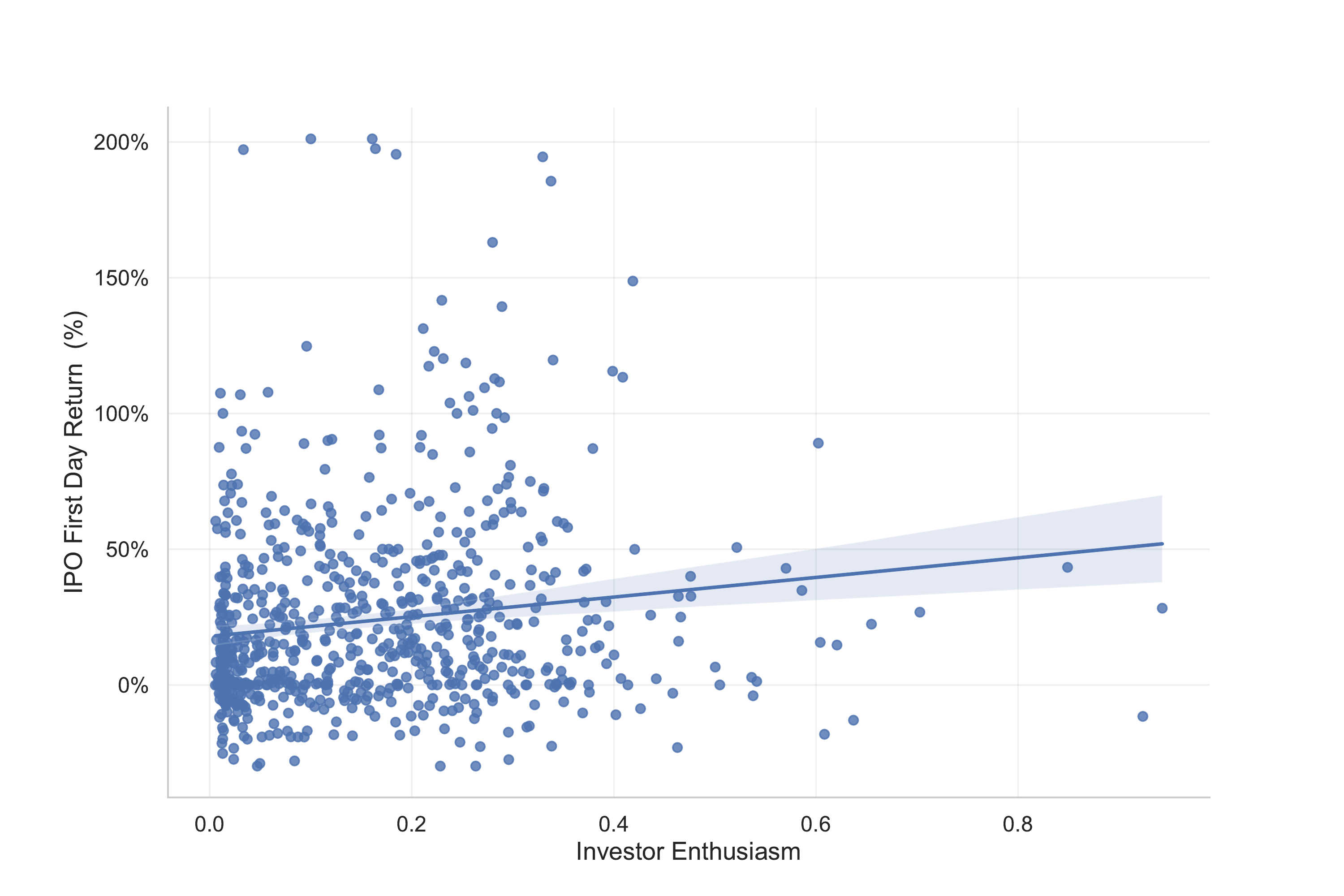}}
    \caption{Investor Enthusiasm and First Day IPO Returns}\label{fig:ipo_firstday}
\end{center}
\begin{flushleft}
 \footnotesize{Notes: Plots based on StockTwits data. }
\end{flushleft}
\end{figure}

In Panel A of Figure \ref{fig:ipo_firstday}, I present a graphical illustration that explores the relationship between investor enthusiasm prior to an IPO and the ensuing first-day IPO return. Panel (a) reveals that IPOs associated with investor enthusiasm falling below the median value manifest an average first-day return of 17.59\%. Conversely, those IPOs where the investor enthusiasm surpasses the median level boast a significantly higher average first-day return, coming in at 29.73\%. Panel (b) takes a more analytical approach, probing the relationship through a linear regression. It portrays a positive linear trajectory between enthusiasm and first-day returns.

To further explore the relationship between investor enthusiasm and IPO returns, I conducted cross-sectional regressions, as presented in Table \ref{tab:pre_market_ipo}. Column (1) shows the regression of first-day IPO returns on investor enthusiasm, while in Column (2), I expanded the model to include other potential predictors such as the log of offering size, industry return, the first day return of recently conducted IPOs, and Fama-French five factors. Column (3) adds industry and year fixed effects and Column (4) restricts the analysis to IPOs with a minimum of 10 StockTwits posts. The findings indicate that pre-IPO investor enthusiasm has a statistically significant effect at the 1 percent level across all regressions. In addition, Column (1) indicates that pre-IPO emotions can account for around 2\% of the variance observed in first day returns ($R^2$). Moreover, a standard deviation increase in pre-IPO enthusiasm translates into a 14.30\% standard deviation increase in first-day returns. These results underscore the importance of pre-IPO investor enthusiasm in shaping first-day IPO returns. The contrast between Columns (2) and (3) indicates that year and industry fixed effects hold greater significance than the average first-day return of recent IPOs and the offering size. However, the point estimate on enthusiasm remains largely consistent. Last, column (4) shows that the impact of pre-IPO investor enthusiasm on first-day IPO returns is even more pronounced for firms with higher levels of online engagement. This suggests that the level of attention and buzz generated by pre-IPO investors can play a critical role in shaping market expectations and driving initial trading activity.

\begin{sidewaystable}[htbp]\centering
\footnotesize
\begin{threeparttable}
\def\sym#1{\ifmmode^{#1}\else\(^{#1}\)\fi}
\caption{Investor Enthusiasm and IPO Returns \label{tab:pre_market_ipo}}
\begin{tabular*}{\hsize}{@{\hskip\tabcolsep\extracolsep\fill}l*{7}{c}}
\hline\hline
                    &\multicolumn{1}{c}{(1)}&\multicolumn{1}{c}{(2)}&\multicolumn{1}{c}{(3)}&\multicolumn{1}{c}{(4)}&\multicolumn{1}{c}{(5)}&\multicolumn{1}{c}{(6)}&\multicolumn{1}{c}{(7)}\\
                    &\multicolumn{4}{c}{First Day Return}&\multicolumn{3}{c}{12-Month Industry Adjusted Return}\\
& & & & & & &\\[\dimexpr-\normalbaselineskip+2pt]
\hline
& & & & & & &\\[\dimexpr-\normalbaselineskip+2pt]
Enthusiasm               &      0.3614\sym{***}&      0.3128\sym{***}&      0.3068\sym{***}&      0.7829\sym{***}&     -0.4910\sym{***}&     -0.4552\sym{***}&     -0.8292\sym{**} \\
                    &    (0.0935)         &    (0.0918)         &    (0.0936)         &    (0.2794)         &    (0.1507)         &    (0.1703)         &    (0.3784)         \\
& & & & & & &\\[\dimexpr-\normalbaselineskip+2pt]
First Day Return  &                     &                     &                     &                     &     -0.2288\sym{**} &     -0.2195\sym{*}  &     -0.3653\sym{**} \\
                    &                     &                     &                     &                     &    (0.1042)         &    (0.1168)         &    (0.1572)         \\
& & & & & & &\\[\dimexpr-\normalbaselineskip+2pt]
Enthusiasm $\times$ First Day Return              &                     &                     &                     &                     &      0.6693\sym{*}  &      0.6115         &      1.1550\sym{*}  \\
                    &                     &                     &                     &                     &    (0.3980)         &    (0.4366)         &    (0.6227)         \\
& & & & & & &\\[\dimexpr-\normalbaselineskip+2pt]
First Day Return$_{-90, -1}$ &                     &      0.4044\sym{***}&      0.1045         &      0.1740         &     -0.2858\sym{***}&     -0.0600         &      0.1554         \\
                    &                     &    (0.0889)         &    (0.1100)         &    (0.1414)         &    (0.1048)         &    (0.1362)         &    (0.1774)         \\
& & & & & & &\\[\dimexpr-\normalbaselineskip+2pt]
Offer Amount        &                     &      0.0226\sym{**} &      0.0186         &     -0.0121         &      0.0090         &      0.0096         &      0.0022         \\
                    &                     &    (0.0102)         &    (0.0126)         &    (0.0206)         &    (0.0147)         &    (0.0177)         &    (0.0252)         \\
& & & & & & &\\[\dimexpr-\normalbaselineskip+2pt]
Constant            &      0.1790\sym{***}&     -0.3444\sym{*}  &     -0.3090         &      0.2286         &     -0.0412         &     -0.0993         &     -0.2781         \\
                    &    (0.0172)         &    (0.1919)         &    (0.2615)         &    (0.4441)         &    (0.2850)         &    (0.3880)         &    (0.5413)         \\ \hline
& & & & & & &\\[\dimexpr-\normalbaselineskip+2pt]
Controls   & & X & X & X & X & X & X \\
Fixed Effects   & & & X & X & & X & X   \\
$\geq$ 10 Posts            &                     &                  &                   &     X              &                    &                  &           X   \\
Observations        &    745        &    742        &    742        &    398        &    742        &    742        &    398        \\
Standardized Effects &    0.1429&    0.1236&    0.1213&   0.2063&   -0.1274&   -0.1181&   -0.1519 \\
$R^2$               &      0.0204         &      0.0821         &      0.2129         &      0.2136         &      0.0293         &      0.0941         &      0.1508         \\
\hline\hline
\end{tabular*}
\begin{tablenotes}
\footnotesize 
\item Notes: This table presents the relationship between investor enthusiasm and two stylized facts regarding initial public offering (IPO) returns. Columns 1-4 depict the first day return, calculated as the difference between the closing and the IPO price, divided by the IPO price. Columns 5-7 illustrate the 12-month industry adjusted return, computed from three months to 12 months post-IPO. I employ the Fama-French 48-industry classification for industry classification. First Day Return$_{-90, -1}$ corresponds to the average first-day return of recent IPOs from 90 days before the IPO up till the day before. Columns (3), (4), (6), and (7) incorporate year and industry fixed effects. I also report the standardized effects of my main independent variable, investor enthusiasm, on my dependent variables. Robust standard errors are reported in parentheses. \sym{*} \(p<0.10\), \sym{**} \(p<0.05\), \sym{***} \(p<0.01\). Continuous variables are winsorized at the 0.5\% and 99.5\% levels to mitigate the impact of outliers.
\end{tablenotes}
\end{threeparttable}
\end{sidewaystable}

\begin{figure}
\begin{center}
    \subfloat[]{\includegraphics[scale=0.45]{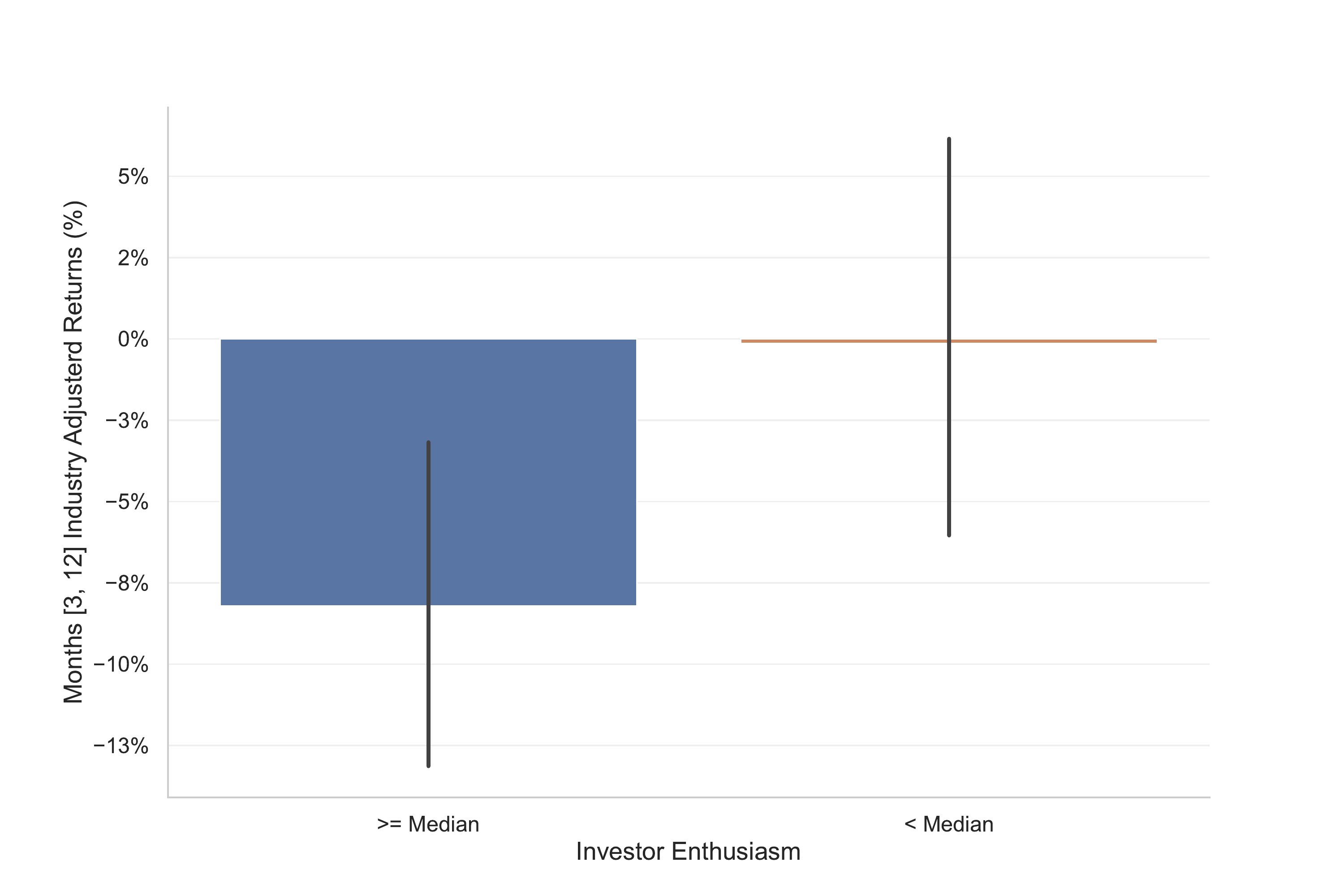}}
    
    \subfloat[]{\includegraphics[scale=0.35]{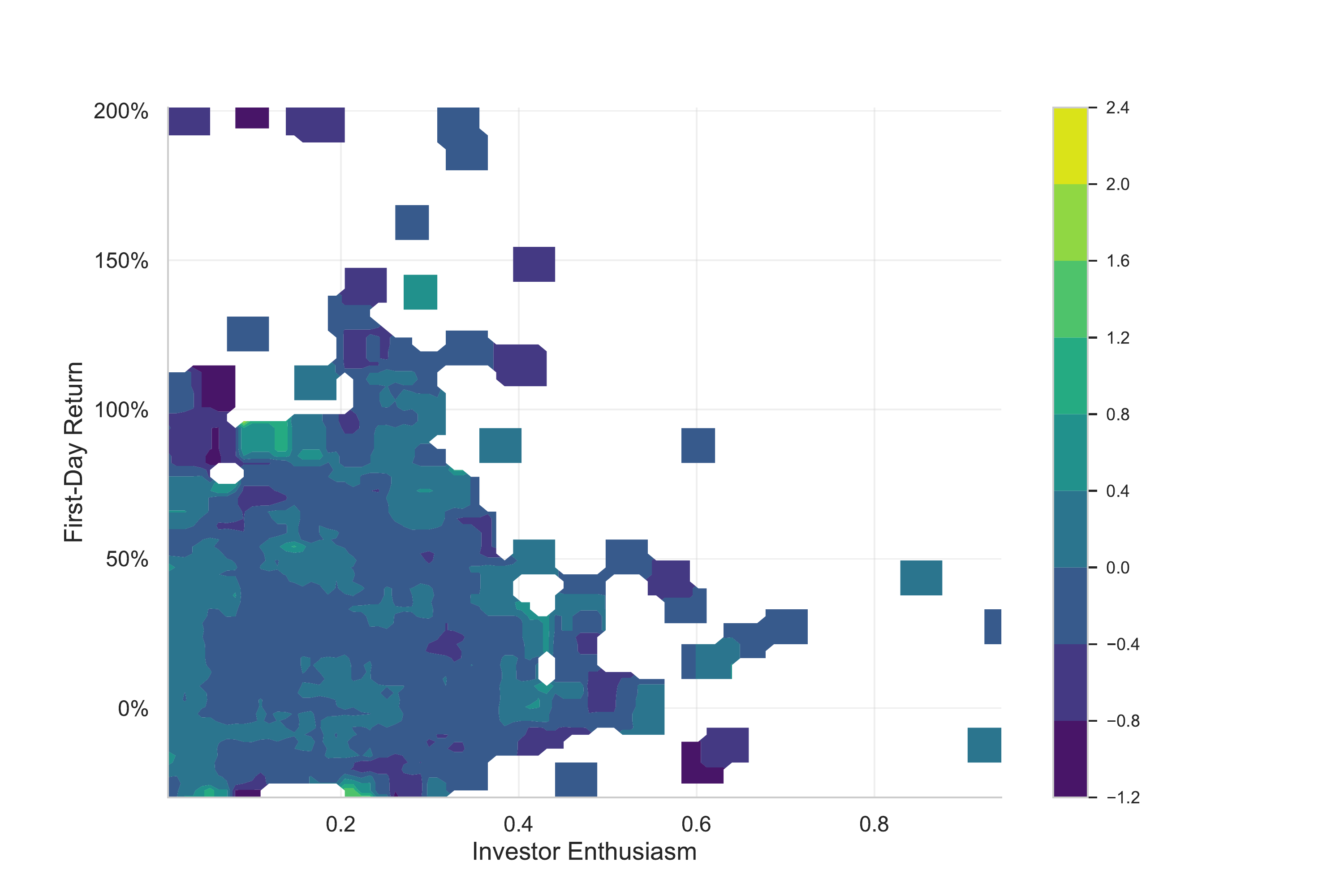}}
    \caption{Investor Enthusiasm and Long Run IPO Returns}\label{fig:ipo_longrun}
\end{center}
\begin{flushleft}
 \footnotesize{Notes: Plots based on StockTwits data. }
\end{flushleft}
\end{figure}

Next, I examine the association between pre-IPO investor enthusiasm and the performance of IPOs over the long term. The primary findings are illustrated in Panel (a) of Figure \ref{fig:ipo_longrun}, which displays the average industry-adjusted returns of IPOs from the end of the third month after the IPO to the end of the twelfth month after the IPO. The first two months following the IPO are excluded since lead underwriters engage in market making and price stabilization during this period (see \cite{ellis2000underwriter}). The plot shows that IPOs with pre-IPO investor enthusiasm below the median level have an average long-run industry-adjusted return of -0.14\%. On the other hand, IPOs with pre-IPO investor enthusiasm above the median level exhibit a significantly lower average long-run industry-adjusted return of -8.22\%.
Furthermore, Panel (b) illustrates the combined impact of first-day returns and pre-IPO investor enthusiasm on long-run industry-adjusted returns. While there's no discernible interaction effect between the two variables, it's crucial to highlight that both higher first-day returns and increased pre-IPO investor enthusiasm generally align with marginally elevated long-run industry-adjusted returns. Consequently, I've incorporated an interaction term between enthusiasm and first-day returns in my regression specifications. The rationale for this is rooted in the theory posited by \cite{ritter2002review}. They contended that the surge of excitement and enthusiasm among retail investors could lead to an overvaluation of an IPO stock, especially evident in its first-day return. Such overvaluation would then pave the way for long-term under-performance as the stock gradually realigns with its intrinsic value. Given this perspective, one would expect that firms with high first-day returns driven by intense investor enthusiasm would underperform in the long run compared to firms that experience high first-day returns without the influence of such heightened enthusiasm, and hence a negative interaction term.

I further refine my analysis using cross-sectional regressions, as shown in Columns (5-7) of Table \ref{tab:pre_market_ipo}, incorporating additional control variables. My findings indicate that both high first-day returns and elevated pre-IPO investor enthusiasm correlate with reduced long-run industry-adjusted returns for IPOs. The observed negative relationships between first-day returns and subsequent long-run performance, coupled with the association between high pre-IPO enthusiasm and these long-run outcomes, suggest that poor performance in the longer term is influenced by both the initial return and investor sentiment. This could imply potential mispricing not fully accounted for by the emotion variable alone, underscoring the significance of first-day returns in this context. This accentuates the crucial role first-day returns play in this dynamic, potentially pointing towards partial price adjustment by underwriters. To quantify, a standard deviation increase in pre-IPO enthusiasm corresponds to a 12.74\% standard deviation decrease in 12-month industry-adjusted returns.

Interestingly, in some specifications, I identify a marginally significant positive interaction term between enthusiasm and first-day returns. This is intriguing because, based on the theory posited by \cite{ritter2002review}, if any, we would anticipate a negative interaction, suggesting that high investor enthusiasm amplifies the overvaluation reflected in first-day returns, leading to greater long-term underperformance. A positive interaction potentially indicates scenarios where high first-day returns and heightened investor enthusiasm might jointly serve as markers of genuine firm potential or other factors that are not straightforwardly captured by our understanding of over-valuation.

Additionally, Columns (5,7) indicate that the impact of pre-IPO investor enthusiasm on long-run returns is notably more pronounced for firms with higher levels of online engagement. Last, the industry and year fixed effects included in Column (6) have minimal influence on my point estimates.

\subsection{Investor Enthusiasm, Information Content, and Returns}

To gain further insights, I refine my analysis by categorizing posts into two main categories and calculating emotions for each. The first category consists of posts containing explicit information about a company's fundamentals and stock price, which I label as ``Finance." The second category includes general chat, which encompasses all messages that are not classified under the ``Finance" category. Moreover, I also distinguish between messages based on whether they provide original information (``Original") or disseminate existing information (``Dissemination"). A message is considered original if it does not contain a hyperlink and is not a retweet of another user's post. These refinements allow me to cast light on the role of emotions in different information environments. 

Table \ref{tab:pre_market_ipo_content} repeats the analysis presented in Table \ref{tab:pre_market_ipo} Column (3) and Column (6), but with emotions assessed separately for messages containing earnings or trading-related information (``Finance'') (Columns 2 and 6) and those conveying other information (``Chat'') (Columns 1 and 5). Next, I also contrast messages containing original information (``Original'') (Columns 3 and 7) and those disseminating existing information (``Dissemination'') (Columns 4 and 8).\footnote{To keep the sample consistent and the point estimates comparable, I restrict the sample to IPOs that have at least one post belonging to each of the four categories.}

\begin{sidewaystable}[htbp]\centering
\footnotesize
\begin{threeparttable}
\def\sym#1{\ifmmode^{#1}\else\(^{#1}\)\fi}
\caption{Investor Enthusiasm, Information Content, and IPO Returns \label{tab:pre_market_ipo_content}}
\begin{tabular*}{\hsize}{@{\hskip\tabcolsep\extracolsep\fill}l*{8}{c}}
\hline\hline
                    &\multicolumn{1}{c}{(1)}&\multicolumn{1}{c}{(2)}&\multicolumn{1}{c}{(3)}&\multicolumn{1}{c}{(4)}&\multicolumn{1}{c}{(5)}&\multicolumn{1}{c}{(6)}&\multicolumn{1}{c}{(7)}&\multicolumn{1}{c}{(8)}\\
Information Content $\rightarrow$ & Chat & Finance & Original & Dissemination & Chat & Finance & Original & Dissemination \\
 Dependent Variable  $\rightarrow$                  &\multicolumn{3}{c}{First Day Return}&\multicolumn{4}{c}{12-Month Industry Adjusted Return}\\
\hline
& & & & & & & & \\[\dimexpr-\normalbaselineskip+2pt]
Enthusiasm                 &      0.2142\sym{**} &      0.3919\sym{***}&      0.0721         &      0.3729\sym{***}&     -0.0383         &     -0.3998\sym{*}  &      0.0548         &     -0.2976         \\
                    &    (0.0897)         &    (0.1308)         &    (0.0716)         &    (0.1402)         &    (0.1546)         &    (0.2099)         &    (0.1788)         &    (0.2220)         \\
& & & & & & & & \\[\dimexpr-\normalbaselineskip+2pt]
Enthusiasm $\times$ First Day Return                  &                     &                     &                     &                     &     -0.3295         &      0.6676         &     -0.6752         &      0.2817         \\
                    &                     &                     &                     &                     &    (0.4129)         &    (0.5295)         &    (0.5208)         &    (0.4262)         \\
& & & & & & & & \\[\dimexpr-\normalbaselineskip+2pt]
First Day Return  &                     &                     &                     &                     &     -0.0045         &     -0.1989         &      0.1026         &     -0.1311         \\
                    &                     &                     &                     &                     &    (0.1392)         &    (0.1246)         &    (0.1846)         &    (0.1049)         \\
& & & & & & & & \\[\dimexpr-\normalbaselineskip+2pt]
First Day Return$_{-90, -1}$&      0.1241         &      0.0709         &      0.1157         &      0.0921         &      0.0521         &      0.0524         &      0.0761         &      0.0653         \\
                    &    (0.1270)         &    (0.1263)         &    (0.1290)         &    (0.1276)         &    (0.1367)         &    (0.1381)         &    (0.1347)         &    (0.1389)         \\
& & & & & & & & \\[\dimexpr-\normalbaselineskip+2pt]
Offer Amount       &      0.0104         &      0.0091         &      0.0124         &      0.0090         &      0.0094         &      0.0106         &      0.0074         &      0.0105         \\
                    &    (0.0167)         &    (0.0170)         &    (0.0168)         &    (0.0167)         &    (0.0200)         &    (0.0197)         &    (0.0204)         &    (0.0201)         \\
& & & & & & & & \\[\dimexpr-\normalbaselineskip+2pt]
Constant            &     -0.2193         &     -0.2272         &     -0.2517         &     -0.2872         &     -0.4730         &     -0.4296         &     -0.4496         &     -0.4079         \\
                    &    (0.3704)         &    (0.3746)         &    (0.3722)         &    (0.3715)         &    (0.4548)         &    (0.4442)         &    (0.4596)         &    (0.4542)         \\
\hline
& & & & & & & & \\[\dimexpr-\normalbaselineskip+2pt]
Controls            &           X         &           X         &           X         &           X         &           X         &           X         &           X         &           X         \\
Observations        &    550        &    550        &    550        &    550        &    550        &    550        &    550        &    550        \\
Standardized Effects  &    0.1021&    0.1383&    0.0377&     0.1350&   -0.0131&   -0.1012&    0.0206&    -0.077\\
$R^2$               &      0.2091         &      0.2138         &      0.2012         &      0.2137         &      0.1013         &      0.1044         &      0.1048         &      0.1019         \\

\hline\hline
\end{tabular*}
\begin{tablenotes}
\footnotesize 
\item Notes: This table presents the relationship between investor emotions, information content and two stylized facts regarding initial public offering (IPO) returns. The first dependent variable is the first day return, which is computed as the difference between the closing and the IPO price, divided by the IPO price. This is shown in Columns 1-4. The second dependent variable is the 12-month industry adjusted return, which is computed from three months after the IPO until 12 months after the IPO (Columns 5-8). Controls, along with year and industry fixed effects are used in each column. The industry classification used is the Fama-French 48-industry classification. I also report the standardized effects of my main independent variable, investor enthusiasm, on my dependent variables. Robust standard errors are reported in parentheses. \sym{*} \(p<0.10\), \sym{**} \(p<0.05\), \sym{***} \(p<0.01\). Continuous variables winsorized at the 0.5\% and 99.5\% level to mitigate the impact of outliers. 
\end{tablenotes}
\end{threeparttable}
\end{sidewaystable}

The analysis presented in Column 2 compared to Column 1 reveals that measures of investor enthusiasm based on messages referencing company fundamentals and earnings generally exhibit more significant point estimates. This outcome suggests that there is valuable financial information being discussed in these chats and these have higher predictive power for IPOs. However, non-negligible information is conveyed in seemingly irrelevant messages, which I capture through my extraction of emotions. For instance, a user may post ":) :)" in response to news events or other posts, indicating their excitement without providing any specific information about the stock. An interesting finding is that posts sharing pre-existing information tend to have a greater impact than original content. This could be due to users referencing online articles, retweeting one another, and other similar activities that generate hype. Last, when assessing long-run industry-adjusted returns, the point estimate for ``Finance" is statistically significant, in contrast to other categories. This suggests that the hype surrounding financial information is predictive of the long-run under-performance.

\subsection{Persistence of IPO Enthusiasm}

By examining the same investors over several IPOs, I uncover trends in their emotional states prior to these IPOs. For this, I employed valence as a yardstick: assigning a ``bearish" label to valence scores below 0, and ``bullish" to scores of 0 and above. Despite the limited sample, I segmented investors into distinct categories based on their consistent bullish or bearish sentiments or leaning. This classification focused on investors who posted for at least 3, 4, or 5 IPOs. To understand the variability in emotions over time, I compute the within-user autocorrelation of their respective valence scores. This metric provides insight into whether an investor's enthusiasm toward a specific IPO is influenced by their enthusiasm toward the prior IPO they interacted with. Formally, this relationship is captured by:

\begin{equation}\label{eq:autocorrelation}
\text{Valence}_{it} =  \alpha_i + \beta_{2} \text{Valence}_{i(t-1)} + \epsilon_{it}
\end{equation}

where $\alpha_i$ is a user fixed effect, capturing their inherent mood trend. $\text{Valence}_{i(t-1)}$ is the valence score from the previous IPO they commented on. Hence, a positive $\beta_2$ implies that investor's emotions about an IPO is generally consistent with how they felt about the previous one. The outcomes are detailed in Table \ref{tab:autocorr}. 

\begin{table}
    \centering
    \caption{Persistence of IPO Enthusiasm \label{tab:autocorr}}
    \begin{threeparttable}
    \begin{tabular}{lcccc} \hline 
        & \multicolumn{3}{c}{Number of IPOs covered} \\ \hline 
        & & & \\[\dimexpr-\normalbaselineskip+2pt] 
         & 3 & 4 & 5 \\ \hline 
         & & & \\[\dimexpr-\normalbaselineskip+2pt] 
        \textbf{Panel A: Valence Pattern} & & & \\
        & & & \\[\dimexpr-\normalbaselineskip+2pt] 
        Always Bullish & 168 & 50 & 22 \\
        Mostly Bullish & 446 & 255 & 200 \\
        Always Bearish & 57 & 13 & 6 \\
        Mostly Bearish & 377 & 214 & 112 \\ \hline 
         & & & \\[\dimexpr-\normalbaselineskip+2pt] 
        \textbf{Panel B: Autocorrelation} & & & \\
        & & & \\[\dimexpr-\normalbaselineskip+2pt] 
        $\beta_2$   &  -0.1697 [0.0143] &  -0.1080 [0.0153] & -0.0713 [0.0160] \\
        $\beta_{2,i} > 0$ & 337 & 174 & 118 \\
        $\beta_{2,i} < 0$ & 646 & 342 & 218 \\ 
        \hline \hline 
    \end{tabular}
        \begin{tablenotes}
            \small
            \item Note: This table delineates the persistence and evolution of investor valence across multiple IPOs. The ``Valence Pattern" panel identifies investors with consistent sentiment directions, categorizing them as either predominantly ``bullish" or ``bearish". The ``Autocorrelation" panel explores how an investor's sentiment toward an IPO might be informed by their valence from the previous IPO they engaged with. Specifically, I compute $\beta_3$ as specified in Equation \ref{eq:autocorrelation} with standard deviations in brackets. To get a sense of heterogeneity, I also compute $beta_{2,i}$, user-level autocorrelation and report the count of users with positive and negative coefficients. Source: StockTwits.
        \end{tablenotes}
    \end{threeparttable}
\end{table}

Panel A offers a detailed breakdown of the valence of investors who have covered a specific minimum number of IPOs. I document an inclination towards bullish sentiments. For instance, the count of investors who are ``Always Bullish" or ``Mostly Bullish" across the different thresholds of IPOs covered is consistently higher than their bearish counterparts. This suggests a general optimism or positive outlook among the investors towards the IPOs they engage with. However, this bullishness isn't universal. The presence of investors in categories ``Always Bearish" or ``Mostly Bearish" showcases the heterogeneity in investor emotions. While a considerable portion of investors display unwavering bullish sentiments, a not-insignificant fraction holds a more cautious or pessimistic outlook. Interestingly, as investors cover more IPOs, the number who consistently maintain an always bullish or always bearish stance sees a reduction, pointing towards a more diverse range of experiences and responses to the IPOs they engage with. 

Panel B sheds light on the process of learning from their experiences with IPOs. The coefficients on $\beta_2$ underscore this phenomenon. For investors who have engaged with 3, 4, or 5 IPOs, the statistically significant negative coefficients at the 1\% level of -0.1697, -0.1080, and -0.0713 respectively indicate a predominant negative, but moderating autocorrelation, suggesting that many of these investors adjust their valence over time. This change in valence is particularly insightful when juxtaposed against the typical trajectory of IPOs. Initially, IPOs often enjoy high first-day returns, sparking a cascade of bullish sentiments. However, the longer-term perspective reveals a different story; when returns are adjusted for industry benchmarks, these IPOs frequently under-perform. The negative autocorrelation possibly signals that investors, having observed this pattern recurrently, grow more circumspect or even skeptical about forthcoming IPOs. They might be undergoing a process of ``learning," wherein they grow less swayed by the allure of immediate high returns and more cautious given the historical trend of lackluster long-term performances.

\subsection{Post-IPO Emotions and Investor Engagement}

\begin{figure}[htbp] 

\begin{center}
	\subfloat[]{\includegraphics[scale=0.45]{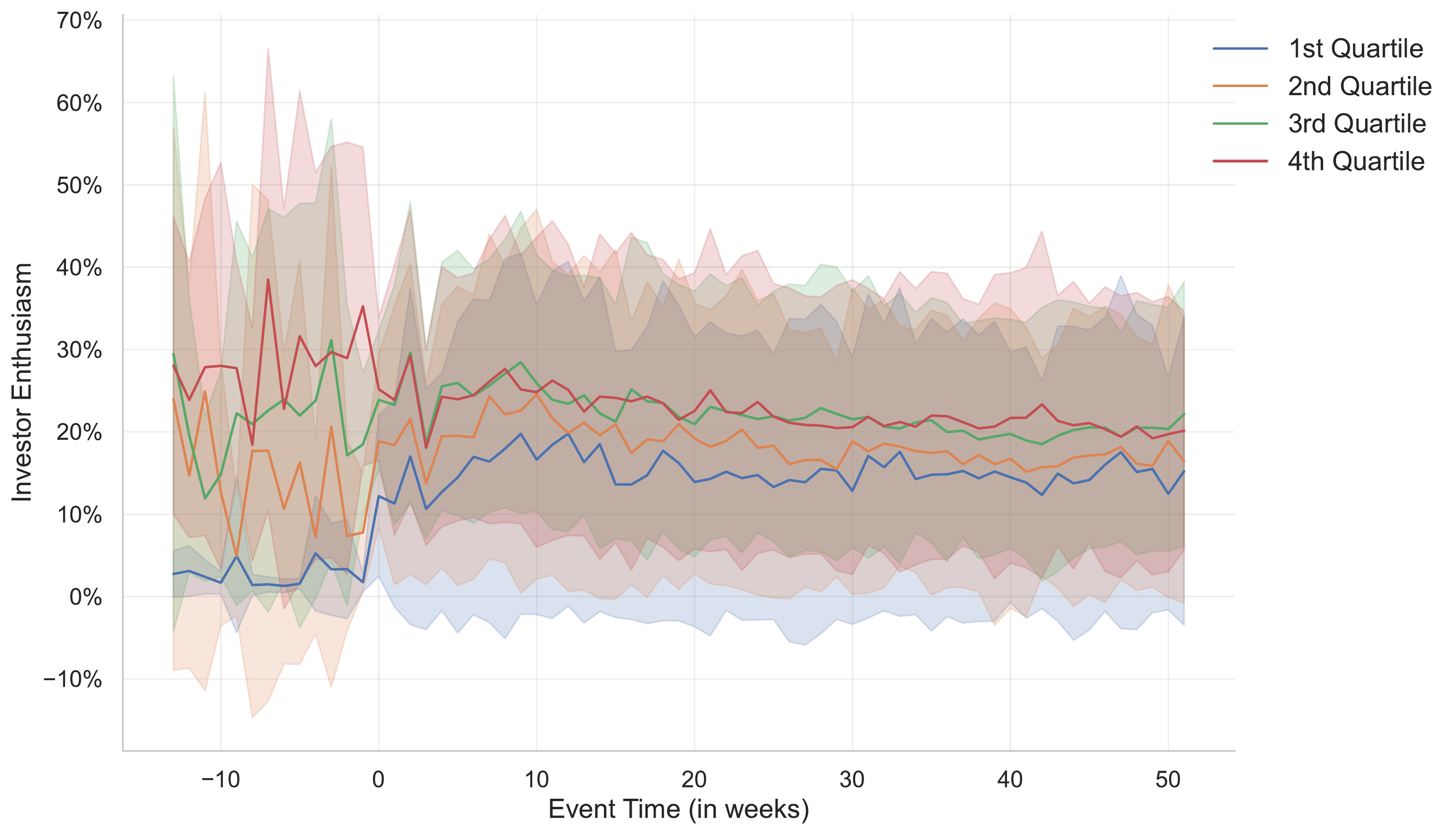}}
 
    \subfloat[]{\includegraphics[scale=0.45]{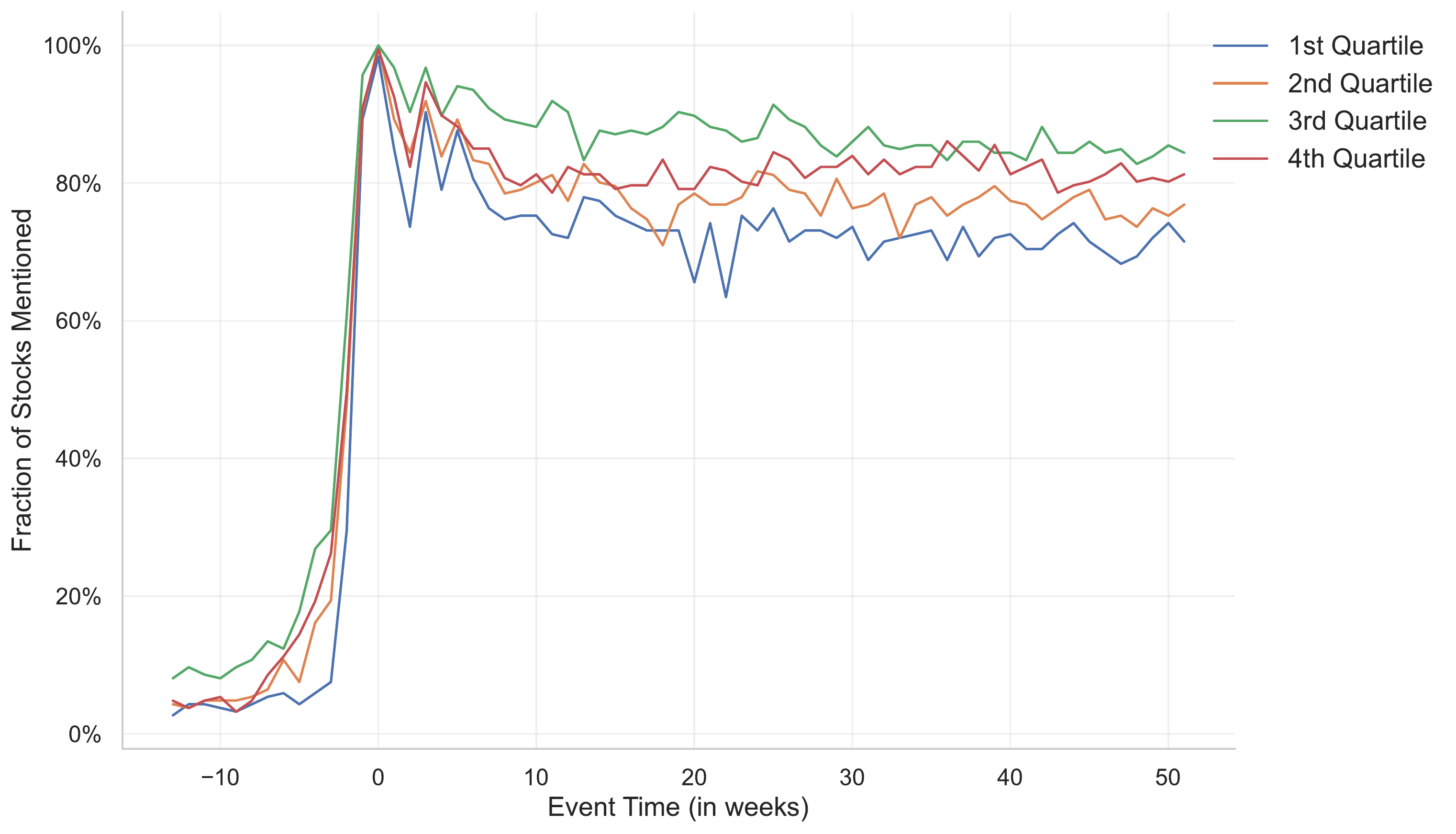}}

	\caption{StockTwits's Coverage of IPOs}\label{fig:ipo_evolution}
\end{center}
 
\begin{flushleft}
	\footnotesize{Notes: Quartiles based on average pre-IPO enthusiasm. Week 0 corresponds to the week of the IPO.}
\end{flushleft}

\end{figure}

While the primary emphasis of this paper is on emotional states leading up to the IPO, the post-IPO enthusiasm trajectory warrants an exploration. Specifically, how does the enthusiasm of investors change following the IPO event? To investigate into this, I segment pre-IPO investor enthusiasm into quartiles, with the 1st quartile indicating the least enthusiastic segment. Using this stratification, in Panel (a) of Figure \ref{fig:ipo_evolution}, I plot the average enthusiasm and its associated standard deviations weekly, starting from 13 weeks before the IPO and extending up to a year afterwards. The figure illustrates some convergence in post-IPO enthusiasm across the quartiles. While there's a visible trend of enthusiasm aligning, it appears to stabilize distinctly for different quartiles: around 20\% for the 3rd and 4th quartiles, and closer to 15\% for the 1st and 2nd quartiles. This differential stabilization could imply that while the initial disparity in enthusiasm might narrow down post-IPO, inherent differences based on pre-IPO perceptions and enthusiasm persist. For the 3rd and 4th quartiles, the sustained higher enthusiasm might indicate a continued optimistic outlook or satisfaction with the IPO's performance. This observation is particularly striking because, as I highlighted in the primary findings, these firms under-perform relative to the industry from 3 months to 12 months post-IPO. This mismatch between sentiment and actual performance underscores the dynamics of investor behavior and sentiment and raises questions about the factors influencing this persistent enthusiasm despite seemingly subpar performance. In contrast, the 1st and 2nd quartiles, though they increase in enthusiasm post-IPO, still remain relatively more cautious or reserved, potentially indicating lingering uncertainties or moderated expectations about the firm's future prospects.

Second, to gauge the continuity of investor engagement, I plot the proportion of IPOs within each quartile that are being discussed on a weekly basis. This offers insights into whether investors, especially those initially exuberant, maintain their discourse momentum or gradually recede from discussions concerning the recent IPO. Panel (b) of Figure \ref{fig:ipo_evolution} visually represents this dynamic. While it's evident that a significant portion of IPO discussions take place in the week leading up to the IPO, the discourse, interestingly, does not wane substantially post-IPO. There is, of course, a dip compared to the week preceding the IPO, but the level of engagement remains high. Firms in the 3rd and 4th quartiles consistently witness a higher proportion of mentions, plateauing at slightly above 80\%. In contrast, firms in the 1st and 2nd quartiles settle at around 71\% and 77\%, respectively. Interestingly, for firms in the 1st and 2nd quartiles post-IPO, investors engage with a smaller fraction of firms (extensive margin) and exhibit a lower average enthusiasm in their discussions (intensive margin) throughout the year following the IPO.

\section{Secondary Findings}\label{sec:sensitivity}

Building upon my initial observations, I first probe the influence of negative emotions, using valence from Equation \ref{eq:emotion}. I then consider investor heterogeneity in aspects like trading approach, experience, and holding period. Concluding this section, I perform a sensitivity analysis to assess potential effects from outliers, language model choice, weighting, and data sources.

\subsection{Valence and IPO Pricing}

Using Valence allows me to discern the extent to which negative emotions influence the prediction. In particular, it helps ascertain whether the prediction is predominantly driven by the sheer buzz and intensity of enthusiasm or if the balance between positive and negative emotions plays a significant role. To do so, I employ ``Valence" as a key independent variable:

\begin{equation}\label{eq:valence}
Y_{it} = \alpha + \beta_3 \text{Valence}_{it} + \zeta'X_{it} + \epsilon_{it}
\end{equation}

Table \ref{tab:pre_market_ipo_valence} offers a comparison by displaying cross-sectional regressions analogous to those in Table \ref{tab:pre_market_ipo}. Notably, the results indicate a more subdued impact and consistently lower $R^2$ values. Take Column (1) as an example. Here, a standard deviation increase in valence corresponds to a 12.75\% increase in the standard deviation of first-day returns. In contrast, when examining enthusiasm in the same specification, the effect is more pronounced at 14.30\%. Another telling comparison emerges from Column (5). In this instance, a one standard deviation uptick in valence leads to a decline of 6.36\% in the standard deviation of industry-adjusted returns. For enthusiasm, this is steeper, standing at 12.74\%. The reduced influence of valence becomes even more evident when additional controls are introduced. For instance, once we account for industry and year fixed effects in Column (6), the valence coefficient loses its statistical significance.

Consequently, the comparative analysis between valence and enthusiasm highlights that IPO predictions are more influenced by the intensity of enthusiasm rather than the balance between positive and negative emotions. This suggests that the market's initial reactions to IPOs are more swayed by the sheer magnitude of excitement than by emotions.

\begin{sidewaystable}[htbp]\centering
\footnotesize
\begin{threeparttable}
\def\sym#1{\ifmmode^{#1}\else\(^{#1}\)\fi}
\caption{Investor Valence and IPO Returns \label{tab:pre_market_ipo_valence}}
\begin{tabular*}{\hsize}{@{\hskip\tabcolsep\extracolsep\fill}l*{7}{c}}
\hline\hline
                    &\multicolumn{1}{c}{(1)}&\multicolumn{1}{c}{(2)}&\multicolumn{1}{c}{(3)}&\multicolumn{1}{c}{(4)}&\multicolumn{1}{c}{(5)}&\multicolumn{1}{c}{(6)}&\multicolumn{1}{c}{(7)}\\
                    &\multicolumn{4}{c}{First Day Return}&\multicolumn{3}{c}{12-Month Industry Adjusted Return}\\
& & & & & & &\\[\dimexpr-\normalbaselineskip+2pt]
\hline
& & & & & & &\\[\dimexpr-\normalbaselineskip+2pt]
Valence             &      0.2944\sym{***}&      0.2685\sym{***}&      0.2499\sym{***}&      0.7513\sym{***}&     -0.2254\sym{*}  &     -0.1665         &     -0.6341\sym{*}  \\
                    &    (0.0730)         &    (0.0725)         &    (0.0695)         &    (0.2634)         &    (0.1291)         &    (0.1360)         &    (0.3443)         \\
& & & & & & &\\[\dimexpr-\normalbaselineskip+2pt]
First Day Return  &                     &                     &                     &                     &     -0.1224         &     -0.1248         &     -0.2467\sym{**} \\
                    &                     &                     &                     &                     &    (0.0749)         &    (0.0829)         &    (0.1177)         \\
& & & & & & &\\[\dimexpr-\normalbaselineskip+2pt]
Valence $\times$ First Day Return            &                     &                     &                     &                     &      0.1466         &      0.1567         &      1.0385         \\
                    &                     &                     &                     &                     &    (0.4082)         &    (0.4338)         &    (0.7168)         \\
& & & & & & &\\[\dimexpr-\normalbaselineskip+2pt]
First Day Return$_{-90, -1}$&                     &      0.4027\sym{***}&      0.1157         &      0.2042         &     -0.2865\sym{***}&     -0.0642         &      0.1449         \\
                    &                     &    (0.0894)         &    (0.1104)         &    (0.1449)         &    (0.1056)         &    (0.1370)         &    (0.1776)         \\
& & & & & & &\\[\dimexpr-\normalbaselineskip+2pt]
Offer Amount        &                     &      0.0263\sym{***}&      0.0234\sym{*}  &     -0.0015         &      0.0044         &      0.0043         &     -0.0055         \\
                    &                     &    (0.0100)         &    (0.0124)         &    (0.0203)         &    (0.0151)         &    (0.0178)         &    (0.0243)         \\

& & & & & & &\\[\dimexpr-\normalbaselineskip+2pt]
Constant            &      0.2159\sym{***}&     -0.3841\sym{**} &     -0.3808         &      0.1484         &     -0.0138         &     -0.0480         &     -0.2523         \\
                    &    (0.0128)         &    (0.1897)         &    (0.2565)         &    (0.4447)         &    (0.2905)         &    (0.3938)         &    (0.5406)         \\ \hline
& & & & & & &\\[\dimexpr-\normalbaselineskip+2pt]
Controls   & & X & X & X & X & X & X \\
Fixed Effects   & & & X & X & & X & X   \\
$\geq$ 10 Posts            &                     &                  &                   &     X              &                    &                  &           X   \\
Observations        &    745        &    742        &    742        &    398        &    742        &    742        &    398        \\
Standardized Effects  &    0.1274&    0.1154&    0.1074&     0.1678&   -0.0636&   -0.04670&   -0.0985\\
$R^2$               &      0.0162         &      0.0804         &      0.2116         &      0.2102         &      0.0216         &      0.0878         &      0.1458         \\
\hline\hline
\end{tabular*}
\begin{tablenotes}
\footnotesize 
\item Notes: This table presents the relationship between investor valence and two stylized facts regarding initial public offering (IPO) returns. Columns 1-4 depict the first day return, calculated as the difference between the closing and the IPO price, divided by the IPO price. Columns 5-7 illustrate the 12-month industry adjusted return, computed from three months to 12 months post-IPO. I employ the Fama-French 48-industry classification for industry classification. First Day Return$_{-90, -1}$ corresponds to the average first-day return of recent IPOs from 90 days before the IPO up till the day before. Columns (3), (4), (6), and (7) incorporate year and industry fixed effects. I also report the standardized effects of my main independent variable, valence, on my dependent variables. Robust standard errors are reported in parentheses. \sym{*} \(p<0.10\), \sym{**} \(p<0.05\), \sym{***} \(p<0.01\). Continuous variables are winsorized at the 0.5\% and 99.5\% levels to mitigate the impact of outliers.
\end{tablenotes}
\end{threeparttable}
\end{sidewaystable}

\subsection{Emotion Heterogeneity and IPO Pricing}

\cite{hong2004groups} demonstrate that a heterogeneous group of informed decision-makers consistently outperforms a more homogenous group, even if the latter comprises individuals with superior abilities. To explore this, I categorize the messages I receive based on the traders' investment durations (long-term vs. short-term), their trading methodologies (value vs. technical), and their trading expertise (novice and intermediate vs. professional). I report heterogeneity across user types in Table \begin{sidewaystable}[htbp]\centering
\footnotesize
\begin{threeparttable}
\def\sym#1{\ifmmode^{#1}\else\(^{#1}\)\fi}
\caption{Emotions, Investor Heterogeneity, and IPO Returns \label{tab:pre_market_ipo_het}}
\begin{tabular*}{\hsize}{@{\hskip\tabcolsep\extracolsep\fill}l*{6}{c}}
\hline\hline
                    &\multicolumn{1}{c}{(1)}&\multicolumn{1}{c}{(2)}&\multicolumn{1}{c}{(3)}&\multicolumn{1}{c}{(4)}&\multicolumn{1}{c}{(5)}&\multicolumn{1}{c}{(6)}\\
& & & & & &\\[\dimexpr-\normalbaselineskip+2pt]
                    & \multicolumn{2}{c}{Horizon}& \multicolumn{2}{c}{Experience}& \multicolumn{2}{c}{Approach} \\
& & & & & &\\[\dimexpr-\normalbaselineskip+2pt]
Investor Type $\rightarrow$ & Short-Term & Long-Term & Novice \& Intermediate & Professional & Fundamental & Technical \\
& & & & & &\\[\dimexpr-\normalbaselineskip+2pt]
\hline
& & & & & &\\[\dimexpr-\normalbaselineskip+2pt]
& & & & & &\\[\dimexpr-\normalbaselineskip+2pt]
\multicolumn{7}{l}{\textbf{Panel A: First Day Return}} \\
& & & & & &\\[\dimexpr-\normalbaselineskip+2pt]
Enthusiasm               &      0.1222         &      0.1796\sym{*}  &      0.1260         &      0.1606         &      0.2677\sym{**} &      0.0036         \\
                    &    (0.1113)         &    (0.0988)         &    (0.0887)         &    (0.1265)         &    (0.1179)         &    (0.1113)         \\
& & & & & &\\[\dimexpr-\normalbaselineskip+2pt]
First Day Return$_{-90, -1}$&      0.1567         &      0.1492         &      0.1576         &      0.1482         &      0.1472         &      0.1623         \\
                    &    (0.1266)         &    (0.1274)         &    (0.1267)         &    (0.1279)         &    (0.1260)         &    (0.1277)         \\
& & & & & &\\[\dimexpr-\normalbaselineskip+2pt]
Offer Amount        &      0.0055         &      0.0057         &      0.0064         &      0.0060         &      0.0022         &      0.0076         \\
                    &    (0.0174)         &    (0.0173)         &    (0.0173)         &    (0.0176)         &    (0.0172)         &    (0.0174)         \\
& & & & & &\\[\dimexpr-\normalbaselineskip+2pt]
Constant            &     -0.1109         &     -0.1094         &     -0.1209         &     -0.1303         &     -0.0670         &     -0.1241         \\
                    &    (0.3710)         &    (0.3692)         &    (0.3707)         &    (0.3702)         &    (0.3666)         &    (0.3708)         \\
\hline
& & & & & &\\[\dimexpr-\normalbaselineskip+2pt]
Observations        &    491        &    491        &    491        &    491        &    491        &    491        \\
$R^2$               &      0.2129         &      0.2156         &      0.2139         &      0.2138         &      0.2210         &      0.2109         \\
\hline
& & & & & &\\[\dimexpr-\normalbaselineskip+2pt]
\multicolumn{7}{l}{\textbf{Panel B: Industry Adjusted Returns}} \\
& & & & & &\\[\dimexpr-\normalbaselineskip+2pt]
Enthusiasm                 &     -0.2619         &     -0.1045         &     -0.1957         &     -0.2747         &     -0.1318         &     -0.3631\sym{**} \\
                    &    (0.1776)         &    (0.1665)         &    (0.1401)         &    (0.2040)         &    (0.2147)         &    (0.1725)         \\
& & & & & &\\[\dimexpr-\normalbaselineskip+2pt]
Enthusiasm $\times$ First Day Return                &      1.1350\sym{**} &     -0.9750\sym{*}  &      0.7464         &      0.1516         &      0.2383         &      0.8345\sym{*}  \\
                    &    (0.5102)         &    (0.5330)         &    (0.4615)         &    (0.5540)         &    (0.4975)         &    (0.4384)         \\
& & & & & &\\[\dimexpr-\normalbaselineskip+2pt]
First Day Return  &     -0.3302\sym{***}&      0.1102         &     -0.2421\sym{**} &     -0.1522         &     -0.1636         &     -0.2782\sym{**} \\
                    &    (0.1271)         &    (0.1753)         &    (0.1143)         &    (0.1630)         &    (0.1419)         &    (0.1233)         \\
& & & & & &\\[\dimexpr-\normalbaselineskip+2pt]
First Day Return$_{-90, -1}$&     -0.0646         &     -0.0193         &     -0.0580         &     -0.0404         &     -0.0508         &     -0.0552         \\
                    &    (0.1454)         &    (0.1538)         &    (0.1465)         &    (0.1528)         &    (0.1479)         &    (0.1481)         \\
& & & & & &\\[\dimexpr-\normalbaselineskip+2pt]
Offer Amount       &     -0.0174         &     -0.0152         &     -0.0180         &     -0.0151         &     -0.0158         &     -0.0149         \\
                    &    (0.0220)         &    (0.0219)         &    (0.0216)         &    (0.0221)         &    (0.0215)         &    (0.0219)         \\
& & & & & &\\[\dimexpr-\normalbaselineskip+2pt]
Constant            &      0.0912         &     -0.0190         &      0.0759         &      0.0412         &      0.0212         &      0.0675         \\
                    &    (0.4775)         &    (0.4820)         &    (0.4773)         &    (0.4847)         &    (0.4735)         &    (0.4781)         \\
\hline
& & & & & &\\[\dimexpr-\normalbaselineskip+2pt]
Observations        &    491        &    491        &    491        &    491        &    491        &    491        \\
$R^2$               &      0.1262         &      0.1301         &      0.1211         &      0.1194         &      0.1170         &      0.1245         \\ \hline\hline
\end{tabular*}
\begin{tablenotes}
\footnotesize 
\item Notes: This table presents the relationship between investor emotions, investor types and two stylized facts regarding initial public offering (IPO) returns. In Panel (a) dependent variable is the first day return, which is computed as the difference between the closing and the IPO price, divided by the IPO price. In Panel (b), the dependent variable is the 12-month industry adjusted return, which is computed from three months after the IPO until 12 months after the IPO (Columns 5-8). Controls, along with year and industry fixed effects are used in each column. The industry classification used is the Fama-French 48-industry classification. Robust standard errors are reported in parentheses. \sym{*} \(p<0.10\), \sym{**} \(p<0.05\), \sym{***} \(p<0.01\). Continuous variables winsorized at the 0.5\% and 99.5\% level to mitigate the impact of outliers. 
\end{tablenotes}
\end{threeparttable}
\end{sidewaystable}
. Within this table, Panel A designates first day return as the dependent variable, while Panel B focuses on the 12-month industry adjusted return. Several intriguing patterns emerge from this analysis.

In line with the value of diversity hypothesis, the analysis reveals that emotions from homogeneous groups seem to have diminished predictive power for IPO returns. The link between enthusiasm and IPO returns consistently mirrors findings from Table \ref{tab:pre_market_ipo}, with significance achieved in a quarter of the models. When looking at first-day returns, it's the enthusiasm of long-term and fundamental traders that emerge as significant predictors of these returns. In stark contrast, while the initial zeal of technical traders doesn't substantially influence first-day returns, it foreshadows long-term under-performance. Technical traders tend to emphasize price movements and trading signals over a stock's fundamental value. In the context of IPOs, where firm-level price history is absent, their forecasts might falter, capturing only the hype which later results in mispricings that adjust over time.

\subsection{Sensitivity Analysis}

I conducted several robustness tests to verify the robustness of my findings. First, I increased the winsorization threshold to (1\%, 99\%) to examine whether the trimming of extreme values influences the results. Second, I removed the winsorization procedure entirely. Third, I weighted posts by the logarithm of followers plus one ($1+\log(\text{\# of Followers}+1)$) to account for variations in the user's influence on social media. Fourth, I repeated the analysis using an alternative emotion model to investigate whether the results are consistent across different emotion detection methods. Last, I replicated the analysis using Twitter data to assess the generalizability of the findings across different social media platforms. I select Columns (2-3) and Columns (5-6) of Table \ref{tab:pre_market_ipo}, and present the sensitivity analysis in Table \ref{tab:pre_market_ipo_sensitivity}.  Within this table, Panel A designates first day return as the dependent variable, while Panel B focuses on the 12-month industry adjusted returns.

\subsubsection{Alternative Winsorization}

Columns 1 to 4 indicate that the outcome of the analysis appears to be unaffected by the application of various degrees of winsorization. In other words, the results are consistent across different levels of winsorization, and it does not seem to have a significant impact on the findings of the study.

\subsubsection{Alternative Weighting}

The application of follower weighting by $1 + \log(1+\text{\# of Followers})$ (Columns 5 and 6) does not appear to have a significant effect on the results of the study.

\subsubsection{Alternative Model}

The emotion models employed in this paper were specifically designed to analyze StockTwits data, and hence text in financial contexts. In Columns 7-8, I use a different emotion model trained on non-financial contexts (i.e., generic social media chat). Despite the change in the model, the results of the analysis remained mostly unchanged. However, in Column 7, the level of significance becomes marginally insignificant for investor enthusiasm.

\subsubsection{Alternative Data}

Last, I tested the generalizability across different social media platforms by conducting a replication analysis using Twitter data. The results of this analysis are presented in columns 9-10. When applying the StockTwits model to examine the link between long-term IPO returns and investor enthusiasm in the Twitter data, the previously observed relationship was found to disappear.\footnote{When I remove the interaction term between enthusiasm and first-day returns, the relationship remains insignificant across both specifications.} However, the relationship between first-day returns and investor enthusiasm remained robust. 

\begin{sidewaystable}[htbp]\centering
\scriptsize
\begin{threeparttable}
\def\sym#1{\ifmmode^{#1}\else\(^{#1}\)\fi}
\caption{Sensitivity Analysis: Emotions and IPO Returns \label{tab:pre_market_ipo_sensitivity}}
\begin{tabular*}{\hsize}{@{\hskip\tabcolsep\extracolsep\fill}l*{10}{c}}
\hline\hline
                    &\multicolumn{1}{c}{(1)}&\multicolumn{1}{c}{(2)}&\multicolumn{1}{c}{(3)}&\multicolumn{1}{c}{(4)}&\multicolumn{1}{c}{(5)}&\multicolumn{1}{c}{(6)}&\multicolumn{1}{c}{(7)}&\multicolumn{1}{c}{(8)}&\multicolumn{1}{c}{(9)}&\multicolumn{1}{c}{(10)}\\
Exercise $\rightarrow$ &\multicolumn{2}{c}{Higher Winsorization}&\multicolumn{2}{c}{No Winsorization}&\multicolumn{2}{c}{Follower Weighting} &\multicolumn{2}{c}{Alternative Model}&\multicolumn{2}{c}{Twitter Data}\\
\hline
& & & & & & & & & & \\[\dimexpr-\normalbaselineskip+2pt]
\multicolumn{10}{l}{\textbf{Panel A: First Day Returns}} \\
& & & & & & & & & &\\[\dimexpr-\normalbaselineskip+2pt]
& & & & & & & & \\[\dimexpr-\normalbaselineskip+2pt]
Enthusiasm                 &      0.3095\sym{***}&      0.3015\sym{***}&      0.3116\sym{***}&      0.3066\sym{***}&      0.2636\sym{***}&      0.2836\sym{***}&      0.1827         &      0.2336\sym{*}  &      0.6613\sym{***}&      0.5048\sym{***}\\
                    &    (0.0900)         &    (0.0909)         &    (0.0918)         &    (0.0939)         &    (0.0937)         &    (0.0960)         &    (0.1204)         &    (0.1203)         &    (0.1039)         &    (0.1057)         \\
& & & & & & & & \\[\dimexpr-\normalbaselineskip+2pt]
First Day Return$_{-90, -1}$&      0.3871\sym{***}&      0.1075         &      0.4025\sym{***}&      0.1006         &      0.4079\sym{***}&      0.1059         &      0.4075\sym{***}&      0.1157         &      0.4788\sym{***}&      0.0620         \\
                    &    (0.0850)         &    (0.1091)         &    (0.0891)         &    (0.1108)         &    (0.0892)         &    (0.1105)         &    (0.0899)         &    (0.1110)         &    (0.0853)         &    (0.1095)         \\
& & & & & & & & \\[\dimexpr-\normalbaselineskip+2pt]
Offer Amount        &      0.0236\sym{**} &      0.0202\sym{*}  &      0.0219\sym{**} &      0.0177         &      0.0240\sym{**} &      0.0193         &      0.0266\sym{***}&      0.0222\sym{*}  &     -0.0005         &     -0.0005         \\
                    &    (0.0097)         &    (0.0118)         &    (0.0106)         &    (0.0131)         &    (0.0101)         &    (0.0125)         &    (0.0101)         &    (0.0125)         &    (0.0090)         &    (0.0108)         \\
& & & & & & & & \\[\dimexpr-\normalbaselineskip+2pt]
Constant            &     -0.3619\sym{**} &     -0.3417         &     -0.3298         &     -0.2907         &     -0.3588\sym{*}  &     -0.3316         &     -0.4038\sym{**} &     -0.4069         &      0.0298         &      0.1125         \\
                    &    (0.1811)         &    (0.2449)         &    (0.2003)         &    (0.2719)         &    (0.1911)         &    (0.2632)         &    (0.1913)         &    (0.2606)         &    (0.1648)         &    (0.2110)         \\
\hline
& & & & & & & & \\[\dimexpr-\normalbaselineskip+2pt]
$R^2$               &      0.0841         &      0.2175         &      0.0799         &      0.2097         &      0.0773         &      0.2104         &      0.0698         &      0.2051         &      0.0836         &      0.1733         \\
\hline
& & & & & & & & & & \\[\dimexpr-\normalbaselineskip+2pt]
\multicolumn{10}{l}{\textbf{Panel B: Industry Adjusted Returns}} \\
& & & & & & & & & &\\[\dimexpr-\normalbaselineskip+2pt]
Enthusiasm                 &     -0.4510\sym{***}&     -0.4211\sym{***}&     -0.5312\sym{***}&     -0.4924\sym{***}&     -0.4363\sym{***}&     -0.4161\sym{**} &     -0.6873\sym{***}&     -0.5665\sym{***}&      0.0335         &      0.1725         \\
                    &    (0.1431)         &    (0.1612)         &    (0.1595)         &    (0.1808)         &    (0.1534)         &    (0.1744)         &    (0.1988)         &    (0.2095)         &    (0.1506)         &    (0.1705)         \\
& & & & & & & & \\[\dimexpr-\normalbaselineskip+2pt]
Enthusiasm $\times$ First Day Return                &      0.6413\sym{*}  &      0.5547         &      0.6958\sym{*}  &      0.6662         &      0.5633         &      0.5299         &      0.5294         &      0.5209         &      0.1302         &      0.0537         \\
                    &    (0.3864)         &    (0.4224)         &    (0.4142)         &    (0.4570)         &    (0.3931)         &    (0.4337)         &    (0.5725)         &    (0.6282)         &    (0.2801)         &    (0.3015)         \\
& & & & & & & & \\[\dimexpr-\normalbaselineskip+2pt]
First Day Return  &     -0.2178\sym{**} &     -0.1987\sym{*}  &     -0.2403\sym{**} &     -0.2404\sym{*}  &     -0.1959\sym{**} &     -0.1902\sym{*}  &     -0.2121         &     -0.2087         &     -0.1475\sym{**} &     -0.1107\sym{*}  \\
                    &    (0.0998)         &    (0.1108)         &    (0.1112)         &    (0.1269)         &    (0.0937)         &    (0.1057)         &    (0.1381)         &    (0.1532)         &    (0.0588)         &    (0.0625)         \\
& & & & & & & & \\[\dimexpr-\normalbaselineskip+2pt]
First Day Return$_{-90, -1}$&     -0.2914\sym{***}&     -0.0645         &     -0.2679\sym{**} &     -0.0361         &     -0.2896\sym{***}&     -0.0564         &     -0.2958\sym{***}&     -0.0601         &     -0.2532\sym{*}  &      0.1311         \\
                    &    (0.0957)         &    (0.1286)         &    (0.1120)         &    (0.1411)         &    (0.1050)         &    (0.1367)         &    (0.1058)         &    (0.1364)         &    (0.1369)         &    (0.1701)         \\
& & & & & & & & \\[\dimexpr-\normalbaselineskip+2pt]
Offer Amount        &      0.0091         &      0.0069         &      0.0088         &      0.0119         &      0.0078         &      0.0088         &      0.0079         &      0.0074         &      0.0074         &      0.0042         \\
                    &    (0.0140)         &    (0.0168)         &    (0.0154)         &    (0.0187)         &    (0.0147)         &    (0.0177)         &    (0.0150)         &    (0.0179)         &    (0.0114)         &    (0.0120)         \\
& & & & & & & & \\[\dimexpr-\normalbaselineskip+2pt]
Constant            &     -0.0564         &     -0.0544         &     -0.0294         &     -0.1482         &     -0.0350         &     -0.0765         &      0.0307         &      0.0432         &     -0.1306         &     -0.1758         \\
                    &    (0.2699)         &    (0.3708)         &    (0.3000)         &    (0.4085)         &    (0.2854)         &    (0.3850)         &    (0.2933)         &    (0.4023)         &    (0.2162)         &    (0.2293)         \\ \hline 
& & & & & & & & \\[\dimexpr-\normalbaselineskip+2pt]
$R^2$               &      0.0309         &      0.1013         &      0.0269         &      0.0859         &      0.0264         &      0.0922         &      0.0305         &      0.0931         &      0.0121         &      0.0846         \\ \hline 
& & & & & & & & \\[\dimexpr-\normalbaselineskip+2pt]
Fixed Effects            &                   &           X         &                  &           X         &                    &           X         &                  &           X         &                  &           X         \\
Observations        &    742       &    742       &    742       &    742       &    742       &    742       &    742       &    742       &   1489       &   1489       \\   

\hline\hline
\end{tabular*}
\begin{tablenotes}
\scriptsize 
\item Notes: This table presents the correlation between investor emotions, information content and two stylized facts regarding initial public offering (IPO) returns. The first dependent variable is the first day return, which is computed as the difference between the closing and the IPO price, divided by the IPO price. This is shown in Columns 1-4. The second dependent variable is the 12-month industry adjusted return, which is computed from three months after the IPO (Columns 5-8). The industry classification used is the Fama-French 48-industry classification. Robust standard errors are reported in parentheses. \sym{*} \(p<0.10\), \sym{**} \(p<0.05\), \sym{***} \(p<0.01\). Continuous variables winsorized at the 1\% and 99\% level to mitigate the impact of outliers. 
\end{tablenotes}
\end{threeparttable}
\end{sidewaystable}

In summary, I found that the relationship between first-day IPO returns and investor emotions remained qualitatively consistent across all specifications, but the level of significance varied slightly in a few cases. The same was observed for long-term IPO returns, except for when analyzing Twitter data with the StockTwits model, where the relationship was not found to be present.

\section{Conclusion} \label{sec:conclusion}

I demonstrated that investor emotions can help rationalize two stylized facts about IPO returns. Specifically, I found that investor emotions expressed on social media platforms such as StockTwits and Twitter can play a significant role in explaining the mispricing of IPO stocks. The abundance of information and opinions shared on these platforms can generate hype around certain stocks, leading to irrational buying and selling decisions by investors, which can result in short-term overvaluation of the stock. However, this enthusiasm often results in long-term correction, as the stock's performance fails to meet the inflated expectations. My study shows that high levels of pre-IPO investor enthusiasm lead to significantly higher first-day returns, but this initial enthusiasm may be misplaced, as these IPOs demonstrate lower long-term industry-adjusted returns compared to those with lower pre-IPO investor enthusiasm.

Digging deeper, the research probes into the nuances of pre-IPO enthusiasm at the individual investor level. A distinct pattern emerges: investors who cover multiple IPOs display an overwhelmingly bullish sentiment. Yet, this bullishness disperses as users interact with a wider spectrum of IPOs, revealing a trend of evolving caution, possibly influenced by prior IPO performances. Additionally, from a firm's perspective post-IPO, it's evident that those garnering significant initial enthusiasm continue to magnetize positive sentiment. This continuity of optimism is apparent both at the intensive and the extensive margins.

The findings of this study have profound implications for a spectrum of market participants. For regulators, understanding the ebb and flow of emotions on social media platforms can be instrumental in detecting potential market manipulations or unwarranted hypes, ensuring a more transparent and fair investment environment. Retail investors, often swayed by the prevailing sentiments on these platforms, can benefit from a better understanding of the link between social media emotions and stock performance, enabling more informed investment decisions. Furthermore, institutional investors and firms considering an IPO might leverage insights from social media sentiment analysis to gauge the possible reception of their offerings in the market. In essence, by shedding light on the intricacies between social media-driven emotions and IPO returns, this research could pave the way for more informed policy-making, smarter investment strategies, and a more transparent marketplace.

\pagebreak

\singlespacing
\nocite{*}
\bibliographystyle{plainnat}
\bibliography{bib}

\clearpage

\onehalfspacing

\appendix
\clearpage

\counterwithin{figure}{section}
\counterwithin{table}{section}
\addcontentsline{toc}{section}{Appendices}

\section*{Appendix}\label{app:app}

\section{Variable Definitions and Sources}\label{app:variable_definitions}

Table \ref{tab:variable_definitions} defines the variables used in the analyses. 

\begin{center}
\onehalfspacing
\footnotesize 
\begin{ThreePartTable}

\begin{longtable}[H]{p{1.5in}p{4in}p{0.75in}}
\caption{Variable Definitions}\label{tab:variable_definitions} \\

Variable & Definition & Source \\ \hline 
\endfirsthead

\multicolumn{3}{c}%
{{\bfseries \tablename\ \thetable{} -- continued from previous page}} \\
Variable & Definition & Source \\ \hline 
& & \\[\dimexpr-\normalbaselineskip+5pt]
\endhead

\endlastfoot
IPO First Day Return & Difference between the initial public offering (IPO) price and the closing price of a stock on its IPO date. & NASDAQ \\ 
12-month Industry Adjusted Return & I compute the industry return using the Fama-French industry code, which covers 48 industries. This is done for the period that starts from 3 months after the IPO and extends up to a year after the IPO. I then calculate the difference between the industry return and the return of the IPO firm during the same period. & NASDAQ \\ 
Offer Amount & Total dollar amount of shares offered to the public. Calculated by multiplying the IPO price per share by the number of shares offered to the public.  & NASDAQ \\
1-month Lag Industry Return & Computed using the Fama-French industry code that encompasses 48 industries. Industry performance starting from one month before the IPO and ending the day before the IPO. & COMPUSTAT/CRSP \\ 
Emotion & I utilize a many-to-one deep learning model to classify each message into one of the seven categories (neutral, happy, sad, anger, disgust, surprise, fear), ensuring that the sum of corresponding probabilities adds up to 1. Next, I compute the average probabilities for each emotion separately, by considering messages from 90 days before the IPO until 9:29 am on the first trading day of the IPO.  & Twitter \& StockTwits \\
& & \\ 
 \hline \hline 
\end{longtable}
\end{ThreePartTable}
\end{center}

\clearpage 

\section{Emotional States and IPO Returns}\label{app:emotions}
I break out the emotion measures and run:\footnote{To avoid multi-collinearity, I consider the benchmark group to be neutral.}
\begin{equation}\label{eq:emotion_pretty}
Y_{it} = \alpha + \gamma_1 \text{Happy}_{it} + \gamma_2 \text{Sad}_{it} + \gamma_3 \text{Hate}_{it} + \gamma_4 \text{Surprise}_{it} + \gamma_5 \text{Fear}_{it} + \zeta'X_{it} + \epsilon_{it}
\end{equation}

Table \ref{tab:pre_market_ipo_emotions} presents the outcomes of cross-sectional regressions examining the influence of various emotional states on IPO returns, notably without the inclusion of an interaction term between individual emotional states and first-day returns. Comparing to Table \ref{tab:pre_market_ipo}, $R^2$ increases from 2.04\% to 2.77\% in Column (1), indicating that these negative emotions do contribute to first-day return variations. When evaluating first-day returns, enthusiasm shows similar point estimates as in Table \ref{tab:pre_market_ipo}. However, for long-run industry-adjusted returns, enthusiasm has a diminished point estimate, with fear emerging as the dominant emotion exerting a substantial and statistically significant negative effect. Interestingly, for first-day returns, apathy (which is equated to sadness in this context) demonstrates a pronounced and statistically significant negative impact.

\begin{sidewaystable}[htbp]\centering
\footnotesize
\begin{threeparttable}
\def\sym#1{\ifmmode^{#1}\else\(^{#1}\)\fi}
\caption{Emotions and IPO Returns \label{tab:pre_market_ipo_emotions}}
\begin{tabular*}{\hsize}{@{\hskip\tabcolsep\extracolsep\fill}l*{7}{c}}
\hline\hline
                    &\multicolumn{1}{c}{(1)}&\multicolumn{1}{c}{(2)}&\multicolumn{1}{c}{(3)}&\multicolumn{1}{c}{(4)}&\multicolumn{1}{c}{(5)}&\multicolumn{1}{c}{(6)}&\multicolumn{1}{c}{(7)}\\
                    &\multicolumn{4}{c}{First Day Return}&\multicolumn{3}{c}{12-Month Industry Adjusted Return}\\
& & & & & & &\\[\dimexpr-\normalbaselineskip+2pt]
\hline
& & & & & & &\\[\dimexpr-\normalbaselineskip+2pt]
Enthusiasm               &      0.3485\sym{***}&      0.3083\sym{***}&      0.2958\sym{***}&      0.7487\sym{**} &     -0.3126\sym{**} &     -0.2905\sym{**} &     -0.3458         \\
                    &    (0.0914)         &    (0.0900)         &    (0.0902)         &    (0.3004)         &    (0.1218)         &    (0.1379)         &    (0.3509)         \\
& & & & & & &\\[\dimexpr-\normalbaselineskip+2pt]
Sadness                 &     -1.1203\sym{**} &     -0.8274\sym{*}  &     -1.1740\sym{**} &     -2.7314\sym{**} &      1.3384         &      1.0852         &      1.7777         \\
                    &    (0.5083)         &    (0.4807)         &    (0.5070)         &    (1.3415)         &    (0.9138)         &    (0.9369)         &    (1.9904)         \\
& & & & & & &\\[\dimexpr-\normalbaselineskip+2pt]
Hate                &      0.4960         &      0.3335         &      0.3385         &      1.4138         &      0.1321         &      0.0046         &     -0.0118         \\
                    &    (0.4672)         &    (0.4586)         &    (0.4355)         &    (0.9518)         &    (0.4540)         &    (0.4578)         &    (1.3261)         \\
& & & & & & &\\[\dimexpr-\normalbaselineskip+2pt]
Fear                &     -0.1537         &     -0.1804         &     -0.0565         &     -1.1483         &     -0.8194\sym{***}&     -0.8065\sym{***}&     -0.8186         \\
                    &    (0.2276)         &    (0.2225)         &    (0.2334)         &    (0.8872)         &    (0.2413)         &    (0.2614)         &    (1.1772)         \\
& & & & & & &\\[\dimexpr-\normalbaselineskip+2pt]
Surprise            &      0.2942\sym{*}  &      0.2251         &      0.2710\sym{*}  &      0.9753         &     -0.3246         &     -0.5328\sym{**} &     -0.4004         \\
                    &    (0.1534)         &    (0.1371)         &    (0.1499)         &    (0.6462)         &    (0.2161)         &    (0.2235)         &    (0.7651)         \\
& & & & & & &\\[\dimexpr-\normalbaselineskip+2pt]
First Day Return$_{-90, -1}$&                     &      0.3984\sym{***}&      0.0972         &      0.1860         &     -0.2846\sym{***}&     -0.0156         &      0.2072         \\
                    &                     &    (0.0890)         &    (0.1098)         &    (0.1394)         &    (0.1061)         &    (0.1399)         &    (0.1876)         \\
& & & & & & &\\[\dimexpr-\normalbaselineskip+2pt]
Offer Amount        &                     &      0.0224\sym{**} &      0.0185         &     -0.0181         &      0.0092         &      0.0140         &      0.0014         \\
                    &                     &    (0.0103)         &    (0.0130)         &    (0.0220)         &    (0.0149)         &    (0.0187)         &    (0.0269)         \\
& & & & & & &\\[\dimexpr-\normalbaselineskip+2pt]
First Day Return &                     &                     &                     &                     &     -0.1007\sym{*}  &     -0.0968         &     -0.1414\sym{*}  \\
                    &                     &                     &                     &                     &    (0.0568)         &    (0.0639)         &    (0.0850)         \\
& & & & & & &\\[\dimexpr-\normalbaselineskip+2pt]
Constant            &      0.1801\sym{***}&     -0.3360\sym{*}  &     -0.2986         &      0.5179         &     -0.0441         &     -0.1772         &     -0.3289         \\
                    &    (0.0212)         &    (0.1941)         &    (0.2666)         &    (0.4814)         &    (0.2856)         &    (0.3991)         &    (0.5668)         \\ \hline
& & & & & & &\\[\dimexpr-\normalbaselineskip+2pt]
Controls   & & X & X & X & X & X & X \\
Fixed Effects   & & & X & X & & X & X   \\
$\geq$ 10 Posts            &                     &                  &                   &     X              &                    &                  &           X   \\
Observations        &    745       &    742       &    742       &    398       &    742       &    742       &    398       \\
$R^2$               &      0.0277         &      0.0865         &      0.2183         &      0.2290         &      0.0338         &      0.1004         &      0.1462         \\
\hline\hline
\end{tabular*}
\begin{tablenotes}
\footnotesize 
\item Notes: This table presents the relationship between investor emotions and two stylized facts regarding initial public offering (IPO) returns. Columns 1-4 depict the first day return, calculated as the difference between the closing and the IPO price, divided by the IPO price. Columns 5-7 illustrate the 12-month industry adjusted return, computed from three months to 12 months post-IPO. I employ the Fama-French 48-industry classification for industry classification. First Day Return$_{-90, -1}$ corresponds to the average first-day return of recent IPOs from 90 days before the IPO up till the day before. Columns (3), (4), (6), and (7) incorporate year and industry fixed effects. Robust standard errors are reported in parentheses. \sym{*} \(p<0.10\), \sym{**} \(p<0.05\), \sym{***} \(p<0.01\). Continuous variables are winsorized at the 0.5\% and 99.5\% levels to mitigate the impact of outliers.
\end{tablenotes}
\end{threeparttable}
\end{sidewaystable}

\clearpage 

\section{Social Media Data}\label{app:social_media_data}

\subsection{StockTwits Activity \& Sample Distributions}\label{app:stocktwits_activity}

Table \ref{tab:tweets_by_year} provides a detailed breakdown of StockTwits activity within my sample, categorized by calendar year. A striking trend emerges from this data: StockTwits activity witnessed a significant surge throughout my sample period, soaring from 107 messages in 2010 to a substantial 11,318 in 2021. This progression underscores the mounting influence and adoption of social media platforms during the 2010s decade. Complementing this, my data set also broadened in scope, encompassing an increase in firm coverage from just 4 in 2010 to 86 by 2021.

I also compare the sample distributions of messages and firms by the Fama-French 48-industry groupings with all IPOs during my sample period. These findings are delineated in Table \ref{tab:fama_fench}. Out of 45 industries that have IPOs, my sample encapsulates a comprehensive 40. My sample is heavier on banking and business services but contains less pharmaceutical firms.

\begin{table}[htbp]\centering 
\footnotesize
\caption{Distribution of StockTwits Posts by Calendar Year \label{tab:tweets_by_year}}
\def\sym#1{\ifmmode^{#1}\else\(^{#1}\)\fi}
\begin{tabular}{l*{2}{c}}
\hline\hline
& & \\[\dimexpr-\normalbaselineskip+2pt]
                    &\multicolumn{1}{c}{(1)}&\multicolumn{1}{c}{(2)}\\
                    &\multicolumn{1}{c}{Firm Observations} &\multicolumn{1}{c}{Tweets} \\
\hline
& & \\[\dimexpr-\normalbaselineskip+2pt]
\hline
2010                &           4         &         107         \\
                    &                     &                     \\
& & \\[\dimexpr-\normalbaselineskip+2pt]
2011                &          27         &        2189         \\
                    &                     &                     \\
& & \\[\dimexpr-\normalbaselineskip+2pt]
2012                &          24         &         409         \\
                    &                     &                     \\
& & \\[\dimexpr-\normalbaselineskip+2pt]
2013                &          24         &        1848         \\
                    &                     &                     \\
& & \\[\dimexpr-\normalbaselineskip+2pt]
2014                &         112         &        1892         \\
                    &                     &                     \\
& & \\[\dimexpr-\normalbaselineskip+2pt]
2015                &          85         &        2757         \\
                    &                     &                     \\
& & \\[\dimexpr-\normalbaselineskip+2pt]
2016                &          42         &         751         \\
                    &                     &                     \\
& & \\[\dimexpr-\normalbaselineskip+2pt]
2017                &          68         &        4664         \\
                    &                     &                     \\
& & \\[\dimexpr-\normalbaselineskip+2pt]
2018                &         113         &        3142         \\
                    &                     &                     \\
& & \\[\dimexpr-\normalbaselineskip+2pt]
2019                &          97         &        5603         \\
                    &                     &                     \\
& & \\[\dimexpr-\normalbaselineskip+2pt]
2020                &          63         &        8077         \\
                    &                     &                     \\
& & \\[\dimexpr-\normalbaselineskip+2pt]
2021                &          86         &       11318         \\
                    &                     &                     \\
& & \\[\dimexpr-\normalbaselineskip+2pt]
Total               &         745         &       42757         \\
                    &                     &                     \\
\hline\hline
\end{tabular}
\end{table} 
\begin{table}[htbp]\centering
\begin{threeparttable}
\caption{Distribution of StockTwits Posts Based on Fama-French  48-Industry Classification\label{tab:fama_fench}}
\scriptsize 
\def\sym#1{\ifmmode^{#1}\else\(^{#1}\)\fi}
\begin{tabular}{l*{3}{c}}
\hline\hline
& & & \\[\dimexpr-\normalbaselineskip+2pt]
                    &\multicolumn{1}{c}{(1)}&\multicolumn{1}{c}{(2)}&\multicolumn{1}{c}{(3)}\\
Fama-French industry code (48 industries)                    &    In-Sample IPOs (\%)     &         Tweets (\%) &  All IPOs (\%) \\
\hline
& & & \\[\dimexpr-\normalbaselineskip+2pt]
Agriculture         &        0.02&        0.13&        0.13\\
Food Products       &        0.87&        0.81&        0.76\\
Recreation          &        1.31&        1.21&        0.63\\
Entertainment       &        0.38&        0.81&        0.76\\
Printing and Publishing&        0.00&        0.13&        0.13\\
Consumer Goods      &        1.04&        1.21&        1.01\\
Apparel             &        0.38&        0.27&        0.32\\
Healthcare          &        0.33&        1.21&        1.51\\
Medical Equipment   &        1.62&        5.77&        4.16\\
Pharmaceutical Products&        7.05&       27.38&       28.73\\
Chemicals           &        0.19&        0.81&        0.95\\
Rubber and Plastic Products&        0.02&        0.40&        0.32\\
Construction Materials&        0.13&        0.54&        0.63\\
Construction        &        0.01&        0.27&        0.69\\
Fabricated Products &        0.01&        0.13&        0.06\\
Machinery           &        0.10&        0.67&        1.07\\
Electrical Equipment&        0.27&        0.54&        0.69\\
Automobiles and Trucks&        0.68&        0.54&        0.50\\
Shipbuilding, Railroad Equipment&        0.01&        0.27&        0.25\\
Non-Metallic and Industrial Metal Mining&        0.08&        0.54&        0.32\\
Coal                &        0.01&        0.13&        0.13\\
Petroleum and Natural Gas&        0.34&        2.15&        2.52\\
Utilities           &        0.15&        0.94&        1.20\\
Communication       &        0.06&        0.54&        0.50\\
Personal Services   &        0.17&        0.94&        1.01\\
Business Services   &       47.19&       25.91&       19.72\\
Computers           &        0.45&        1.07&        1.01\\
Electronic Equipment&        1.10&        1.88&        2.77\\
Measuring and Control Equipment&        0.16&        0.27&        0.50\\
Business Supplies   &        0.04&        0.27&        0.19\\
Shipping Containers &        0.03&        0.13&        0.13\\
Transportation      &        5.50&        1.88&        2.46\\
Wholesale           &        0.17&        0.94&        1.58\\
Retail              &        4.35&        4.56&        3.53\\
Restaraunts, Hotels, Motels&        6.11&        1.88&        1.26\\
Banking             &       12.05&        5.23&        5.29\\
Insurance           &        0.57&        1.88&        1.64\\
Real Estate         &        0.29&        0.67&        0.57\\
Trading             &        5.08&        4.03&        6.30\\
Almost Nothing      &        1.68&        1.07&        3.47\\
Candy \& Soda        &            &            &        0.06\\
Beer \& Liquor       &            &            &        0.06\\
Tobacco Products    &            &            &        0.06\\
Steel Works Etc     &            &            &        0.38\\
Precious Metals     &            &            &        0.06\\
\hline
& & & \\[\dimexpr-\normalbaselineskip+2pt]
Observations        &       42757&         745&        1587\\
\hline\hline
\end{tabular}
\begin{tablenotes}
\scriptsize 
\item CRSP sample follows the same basic sample restrictions; active, traded on U.S. exchanges.
\end{tablenotes}
\end{threeparttable}
\end{table}

\subsection{Twitter Data}\label{app:twitter_activity}
Table \ref{tab:summary_stats_ipo_tw} presents the descriptive statistics for the analysis variables, while 
Table \ref{tab:corr_matrix_tw} presents pairwise correlation coefficients among my analysis variables for the Twitter sample. 

\begin{table}[htbp]\centering
\begin{threeparttable}
\footnotesize 
\def\sym#1{\ifmmode^{#1}\else\(^{#1}\)\fi}
\caption{Summary Statistics: IPO Sample, Twitter Data \label{tab:summary_stats_ipo_tw}}
\begin{tabular}{l*{7}{c}}
\hline\hline
& \multicolumn{7}{c}{CRSP/Compustat-Twitter-Nasdaq} \\
\hline
& & & & & & & \\[\dimexpr-\normalbaselineskip+2pt] 
 &    Mean &      Std. Dev. &     1\% &    25\% &     50\% &     75\% &      99\% \\ \hline 
 & & & & & & & \\[\dimexpr-\normalbaselineskip+2pt] 
\multicolumn{8}{l}{\textbf{Social Media Information}} \\
& & & & & & & \\[\dimexpr-\normalbaselineskip+2pt] 
\textcolor{white}{...} Number of Posts & 57.392 & 245.648 & 2 & 5 & 11 & 29 & 1114 \\
\textcolor{white}{...} Number of Users & 32.034 & 144.554 & 1 & 3 & 6 & 13 & 589.880 \\
\textcolor{white}{...} Happy & 0.160 & 0.140 & 0.008 & 0.036 & 0.133 & 0.246 & 0.607 \\
\textcolor{white}{...} Sad & 0.017 & 0.022 & 0.001 & 0.004 & 0.008 & 0.023 & 0.110 \\
\textcolor{white}{...} Hate & 0.024 & 0.039 & 0.001 & 0.004 & 0.010 & 0.033 & 0.136 \\
\textcolor{white}{...} Surprise & 0.050 & 0.066 & 0.002 & 0.006 & 0.024 & 0.078 & 0.337 \\
\textcolor{white}{...} Fear & 0.048 & 0.053 & 0.006 & 0.016 & 0.033 & 0.068 & 0.215 \\
\textcolor{white}{...} Valence & 0.071 & 0.153 & -0.389 & -0.002 & 0.043 & 0.136 & 0.574 \\
& & & & & & & \\[\dimexpr-\normalbaselineskip+2pt] 
\multicolumn{8}{l}{\textbf{IPO \& Financial Information}} \\
 & & & & & & & \\[\dimexpr-\normalbaselineskip+2pt] 
\textcolor{white}{...} First Day Return & 0.237 & 0.353 & -0.244 & 0.001 & 0.138 & 0.381 & 1.567 \\
\textcolor{white}{...} IPO Price & 18.392 & 13.761 & 5 & 14 & 16.500 & 20 & 54.240 \\
\textcolor{white}{...} IPO Shares & 17.431 & 29.937 & 1.141 & 5.750 & 9.100 & 17.250 & 134.500 \\
\textcolor{white}{...} Offer Amount & 423.031 & 1403.342 & 10.660 & 80 & 146.625 & 331.579 & 3450.400 \\
\textcolor{white}{...} Industry Adjusted Return$_{2m, 12m}$ & -0.064 & 0.507 & -0.948 & -0.381 & -0.140 & 0.132 & 1.825 \\
\textcolor{white}{...} Industry Adjusted Return$_{3m, 12m}$ & -0.048 & 0.500 & -0.928 & -0.366 & -0.124 & 0.149 & 2.003 \\
\textcolor{white}{...} Industry Adjusted Return$_{4m, 12m}$ & -0.062 & 0.443 & -0.862 & -0.352 & -0.123 & 0.135 & 1.610 \\
\textcolor{white}{...} 1-month Lag Industry Return & 0.031 & 0.239 & -0.495 & -0.121 & 0.019 & 0.151 & 0.805 \\ \hline 
& & & & & & & \\[\dimexpr-\normalbaselineskip+2pt] 
\textcolor{white}{...} Observations &    &     & &  &  &    &   754 \\ \hline \hline
& & & & & & & \\[\dimexpr-\normalbaselineskip+2pt] 
\multicolumn{8}{l}{\textbf{Panel B: Twitter}} \\
& & & & & & & \\[\dimexpr-\normalbaselineskip+2pt] 
\multicolumn{8}{l}{\textbf{Social Media Information}} \\
& & & & & & & \\[\dimexpr-\normalbaselineskip+2pt] 
\textcolor{white}{...} Number of Posts             & 125.948 & 737.399 & 2 & 16 & 29 & 51 & 1735.120 \\
\textcolor{white}{...} Number of Users         & 69.754 & 416.089 & 2 & 11 & 20 & 32 & 1020.620 \\
\textcolor{white}{...} Happy                 & 0.117 & 0.105 & 0.009 & 0.037 & 0.084 & 0.168 & 0.465 \\
\textcolor{white}{...} Sad                     & 0.012 & 0.015 & 0.001 & 0.004 & 0.007 & 0.015 & 0.068 \\
\textcolor{white}{...} Hate                    & 0.015 & 0.024 & 0.001 & 0.004 & 0.008 & 0.017 & 0.118 \\
\textcolor{white}{...} Surprise              & 0.022 & 0.028 & 0.002 & 0.005 & 0.011 & 0.029 & 0.130 \\
\textcolor{white}{...} Fear                      & 0.043 & 0.036 & 0.007 & 0.020 & 0.035 & 0.055 & 0.151 \\
\textcolor{white}{...} Valence & 0.047	& 0.102	&-0.169	&-0.005	&0.023&	0.081&	0.412 \\
& & & & & & & \\[\dimexpr-\normalbaselineskip+2pt] 
\multicolumn{8}{l}{\textbf{IPO \& Financial Information}} \\
 & & & & & & & \\[\dimexpr-\normalbaselineskip+2pt] 
\textcolor{white}{...} First Day Return      & 0.202 & 0.342 & -0.274 & 0 & 0.103 & 0.307 & 1.633 \\
\textcolor{white}{...} IPO Price               & 17.083 & 11.750 & 5 & 12 & 16 & 19 & 49.110 \\
\textcolor{white}{...} IPO Shares             & 14.984 & 23.516 & 1.110 & 5.500 & 8.427 & 15.665 & 115.158 \\
\textcolor{white}{...} Offer Amount         & 319.781 & 1020.978 & 8.722 & 75 & 131.020 & 271.710 & 2877.530 \\
\textcolor{white}{...} Industry Adjusted Return$_{2m, 12m}$  & -0.079 & 0.490 & -0.940 & -0.389 & -0.141 & 0.118 & 1.831 \\
\textcolor{white}{...} Industry Adjusted Return$_{3m, 12m}$ & -0.070 & 0.473 & -0.893 & -0.362 & -0.127 & 0.115 & 1.994 \\
\textcolor{white}{...} Industry Adjusted Return$_{4m, 12m}$ & -0.069 & 0.422 & -0.857 & -0.337 & -0.110 & 0.125 & 1.610 \\
\textcolor{white}{...} 1-month Lag Industry Return & 0.038 & 0.243 & -0.483 & -0.114 & 0.018 & 0.163 & 0.810 \\ \hline 
& & & & & & & \\[\dimexpr-\normalbaselineskip+2pt] 
\textcolor{white}{...} Observations &    &     & &  &  &    &   1570 \\ \hline \hline
\end{tabular}
   \begin{tablenotes}
     \footnotesize
    \item Notes: Offer Amount and IPO Shares in millions. Emotions from 90 days before the IPO date up till 9.29AM of the IPO date; firms with at least two messages. Return$_{2m, 12m}$ corresponds to IPO firm return from two months after the IPO until a year after the IPO. Return variables are winsorized at the 0.5\% and 99.5\% levels. Twitter data.
   \end{tablenotes}
\end{threeparttable}
\end{table}
 
\begin{sidewaystable}[htbp]\centering
\begin{threeparttable}
\footnotesize
\caption{Correlation Matrix: Twitter Data}\label{tab:corr_matrix_tw}
\def\sym#1{\ifmmode^{#1}\else\(^{#1}\)\fi}
\begin{tabular}{l*{9}{c}}
\hline\hline
                &\multicolumn{9}{c}{}                                                                                                                                                      \\
                &  Valence         &    Happy         &      Sad         &     Hate         & Surprise         &     Fear         &Industry Adjusted Return$_{3,12}$        &Offer Amount       &       \\
\hline
& & & & & & & \\[\dimexpr-\normalbaselineskip+2pt] 
Happy           &     0.83\sym{***}&            &                  &                  &                  &                  &                  &                  &                  \\
& & & & & & & \\[\dimexpr-\normalbaselineskip+2pt] 

Sad             &    -0.22\sym{***}&     0.20\sym{***}&            &                  &                  &                  &                  &                  &                  \\
& & & & & & & \\[\dimexpr-\normalbaselineskip+2pt] 

Hate            &    -0.16\sym{***}&     0.29\sym{***}&     0.42\sym{***}&            &                  &                  &                  &                  &                  \\
& & & & & & & \\[\dimexpr-\normalbaselineskip+2pt] 

Surprise        &     0.07         &     0.29\sym{***}&     0.24\sym{***}&     0.35\sym{***}&            &                  &                  &                  &                  \\
& & & & & & & \\[\dimexpr-\normalbaselineskip+2pt] 

Fear            &    -0.22\sym{***}&     0.31\sym{***}&     0.52\sym{***}&     0.46\sym{***}&     0.32\sym{***}&            &                  &                  &                  \\
& & & & & & & \\[\dimexpr-\normalbaselineskip+2pt] 

Industry Adjusted Return$_{3,12}$&     0.02         &    -0.00         &    -0.04         &    -0.02         &    -0.01         &    -0.03         &              &                  &                  \\
& & & & & & & \\[\dimexpr-\normalbaselineskip+2pt] 

Offer Amount  &     0.11\sym{**} &     0.19\sym{***}&     0.07         &     0.10\sym{**} &     0.14\sym{***}&     0.15\sym{***}&     0.01         &              &                  \\
& & & & & & & \\[\dimexpr-\normalbaselineskip+2pt] 

First Day Return&     0.17\sym{***}&     0.21\sym{***}&     0.02         &     0.04         &     0.09\sym{**} &     0.10\sym{**} &    -0.09\sym{*}  &     0.06         &              \\
& & & & & & & \\[\dimexpr-\normalbaselineskip+2pt] 
\hline
Observations    &     1490         &                  &                  &                  &                  &                  &                  &                  &                  \\

\hline\hline
\end{tabular}
\begin{tablenotes}
\scriptsize 
\item Notes: Pairwise correlation with Bonferroni corrections. Return variables are winsorized at the 0.5\% and 99.5\% levels. \textit{t} statistics in parentheses. \sym{*} \(p<0.05\), \sym{**} \(p<0.01\), \sym{***} \(p<0.001\). Twitter data.
\end{tablenotes}
\end{threeparttable}
\end{sidewaystable}

\subsubsection{Activity \& Sample Distributions}

Table \ref{tab:tweets_by_year_tw} offers an in-depth overview of Twitter activity in my sample, segmented by calendar year. The trend is similar as in StockTwits. The volume of Twitter messages in the dataset dramatically increased, jumping from 107 messages in 2010 to 63,861 in 2021. Concurrently, the range of firm coverage expanded, moving from 34 in 2010 to 273 by the close of 2021.

I also compare the sample distributions of messages and firms by the Fama-French 48-industry groupings with all IPOs during my sample period. These insights are detailed in Table \ref{tab:fama_fench_tw}. Of the 45 industries undergoing IPOs, my sample encompasses each one. Echoing the StockTwits data, the Twitter sample is more concentrated in banking and business services, yet is lighter on pharmaceutical entities.

\begin{table}[htbp]\centering 
\footnotesize
\caption{Distribution of Twitter Posts by Calendar Year \label{tab:tweets_by_year_tw}}
\def\sym#1{\ifmmode^{#1}\else\(^{#1}\)\fi}
\begin{tabular}{l*{2}{c}}
\hline\hline
& & \\[\dimexpr-\normalbaselineskip+2pt]
                    &\multicolumn{1}{c}{(1)}&\multicolumn{1}{c}{(2)}\\
                    &\multicolumn{1}{c}{Firm Observations} &\multicolumn{1}{c}{Tweets} \\
\hline
& & \\[\dimexpr-\normalbaselineskip+2pt]
\hline
2010                &          34         &         295         \\
                    &                     &                     \\
& & \\[\dimexpr-\normalbaselineskip+2pt]
2011                &          54         &        4088         \\
                    &                     &                     \\
& & \\[\dimexpr-\normalbaselineskip+2pt]
2012                &          88         &        2599         \\
                    &                     &                     \\
& & \\[\dimexpr-\normalbaselineskip+2pt]
2013                &         126         &       19830         \\
                    &                     &                     \\
& & \\[\dimexpr-\normalbaselineskip+2pt]
2014                &         187         &        9685         \\
                    &                     &                     \\
& & \\[\dimexpr-\normalbaselineskip+2pt]
2015                &         122         &       11979         \\
                    &                     &                     \\
& & \\[\dimexpr-\normalbaselineskip+2pt]
2016                &          80         &       11620         \\
                    &                     &                     \\
& & \\[\dimexpr-\normalbaselineskip+2pt]
2017                &         124         &       14919         \\
                    &                     &                     \\
& & \\[\dimexpr-\normalbaselineskip+2pt]
2018                &         134         &        8788         \\
                    &                     &                     \\
& & \\[\dimexpr-\normalbaselineskip+2pt]
2019                &         110         &       11686         \\
                    &                     &                     \\
& & \\[\dimexpr-\normalbaselineskip+2pt]
2020                &         158         &       28312         \\
                    &                     &                     \\
& & \\[\dimexpr-\normalbaselineskip+2pt]
2021                &         273         &       63861         \\
                    &                     &                     \\
& & \\[\dimexpr-\normalbaselineskip+2pt]
Total               &        1490         &      187662         \\
                    &                     &                     \\
\hline\hline
\end{tabular}
\end{table} 
\begin{table}[htbp]\centering
\begin{threeparttable}
\caption{Distribution of Twitter Posts Based on Fama-French  48-Industry Classification\label{tab:fama_fench_tw}}
\scriptsize 
\def\sym#1{\ifmmode^{#1}\else\(^{#1}\)\fi}
\begin{tabular}{l*{3}{c}}
\hline\hline
& & & \\[\dimexpr-\normalbaselineskip+2pt]
                    &\multicolumn{1}{c}{(1)}&\multicolumn{1}{c}{(2)}&\multicolumn{1}{c}{(3)}\\
Fama-French industry code (48 industries)                    &    In-Sample IPOs (\%)     &         Tweets (\%) &  All IPOs (\%) \\
\hline
& & & \\[\dimexpr-\normalbaselineskip+2pt]
Agriculture         &        0.03&        0.13&        0.13\\
Food Products       &        0.41&        0.81&        0.76\\
Candy \& Soda        &        0.01&        0.07&        0.06\\
Beer \& Liquor       &        0.03&        0.07&        0.06\\
Tobacco Products    &        0.02&        0.07&        0.06\\
Recreation          &        0.78&        0.60&        0.63\\
Entertainment       &        0.60&        0.81&        0.76\\
Printing and Publishing&        0.01&        0.13&        0.13\\
Consumer Goods      &        2.85&        1.07&        1.01\\
Apparel             &        0.29&        0.34&        0.32\\
Healthcare          &        0.44&        1.54&        1.51\\
Medical Equipment   &        6.21&        4.30&        4.16\\
Pharmaceutical Products&       15.62&       28.93&       28.73\\
Chemicals           &        0.73&        1.01&        0.95\\
Rubber and Plastic Products&        0.05&        0.27&        0.32\\
Construction Materials&        0.11&        0.67&        0.63\\
Construction        &        0.09&        0.67&        0.69\\
Steel Works Etc     &        0.02&        0.34&        0.38\\
Fabricated Products &        0.02&        0.07&        0.06\\
Machinery           &        0.31&        1.07&        1.07\\
Electrical Equipment&        0.20&        0.54&        0.69\\
Automobiles and Trucks&        0.88&        0.47&        0.50\\
Shipbuilding, Railroad Equipment&        0.04&        0.27&        0.25\\
Precious Metals     &        0.03&        0.07&        0.06\\
Non-Metallic and Industrial Metal Mining&        0.09&        0.34&        0.32\\
Coal                &        0.03&        0.13&        0.13\\
Petroleum and Natural Gas&        0.43&        2.42&        2.52\\
Utilities           &        0.23&        1.07&        1.20\\
Communication       &        0.11&        0.40&        0.50\\
Personal Services   &        0.37&        0.94&        1.01\\
Business Services   &       35.15&       20.27&       19.72\\
Computers           &        0.41&        1.07&        1.01\\
Electronic Equipment&        0.96&        2.62&        2.77\\
Measuring and Control Equipment&        0.37&        0.47&        0.50\\
Business Supplies   &        0.03&        0.20&        0.19\\
Shipping Containers &        0.05&        0.13&        0.13\\
Transportation      &        3.07&        2.42&        2.46\\
Wholesale           &        0.44&        1.68&        1.58\\
Retail              &        5.97&        3.69&        3.53\\
Restaraunts, Hotels, Motels&        3.43&        1.28&        1.26\\
Banking             &       10.59&        5.23&        5.29\\
Insurance           &        0.61&        1.61&        1.64\\
Real Estate         &        2.34&        0.54&        0.57\\
Trading             &        4.82&        5.70&        6.30\\
Almost Nothing      &        0.70&        3.49&        3.47\\
\hline
& & & \\[\dimexpr-\normalbaselineskip+2pt]
Observations        &      187662&        1490&        1587\\
\hline\hline
\end{tabular}
\begin{tablenotes}
\scriptsize 
\item CRSP sample follows the same basic sample restrictions; active, traded on U.S. exchanges.
\end{tablenotes}
\end{threeparttable}
\end{table}

\subsubsection{Coverage}

\begin{figure}[htbp] 

\begin{center}
	\subfloat[]{\includegraphics[scale=0.33]{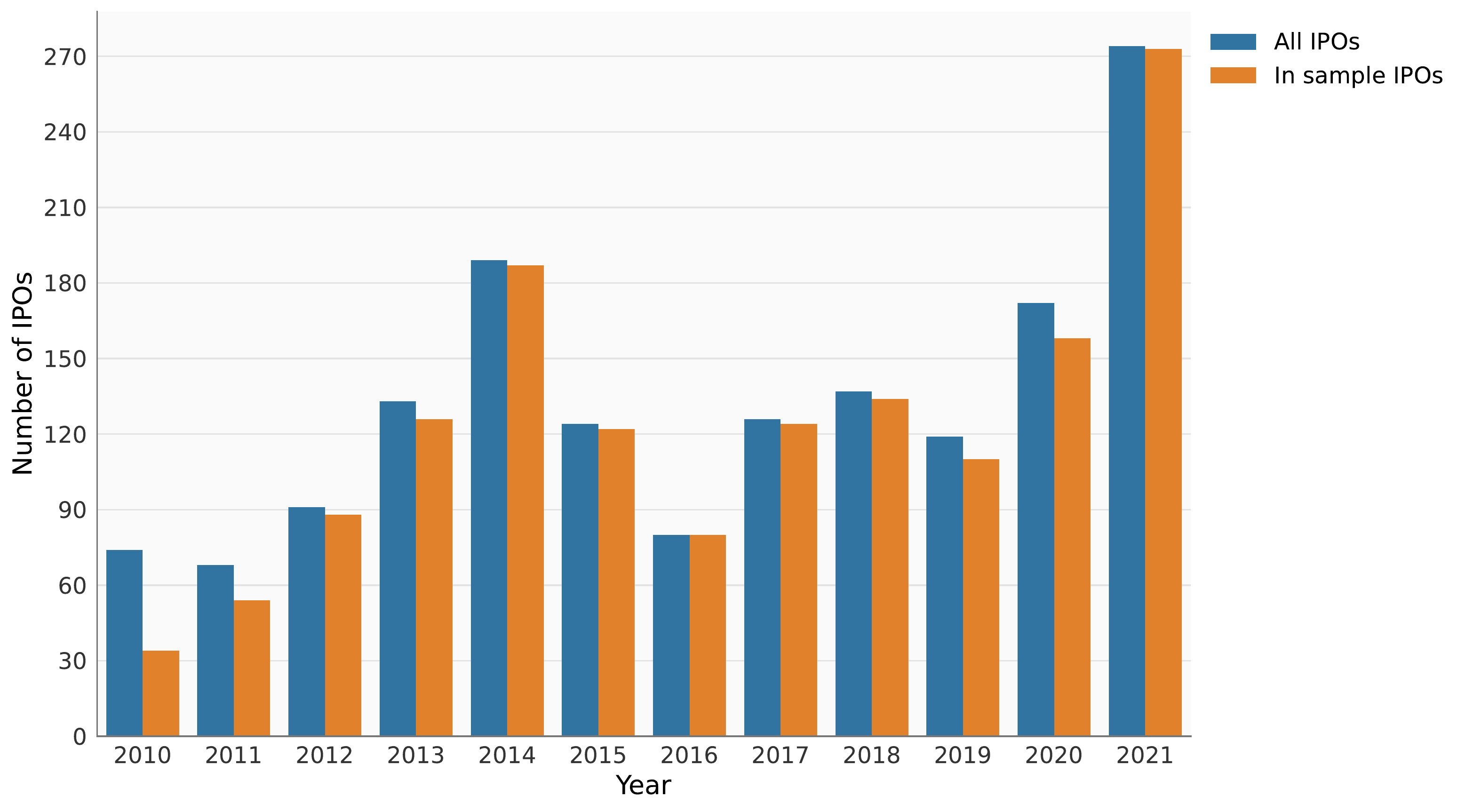}}
 
    \subfloat[]{\includegraphics[scale=0.33]{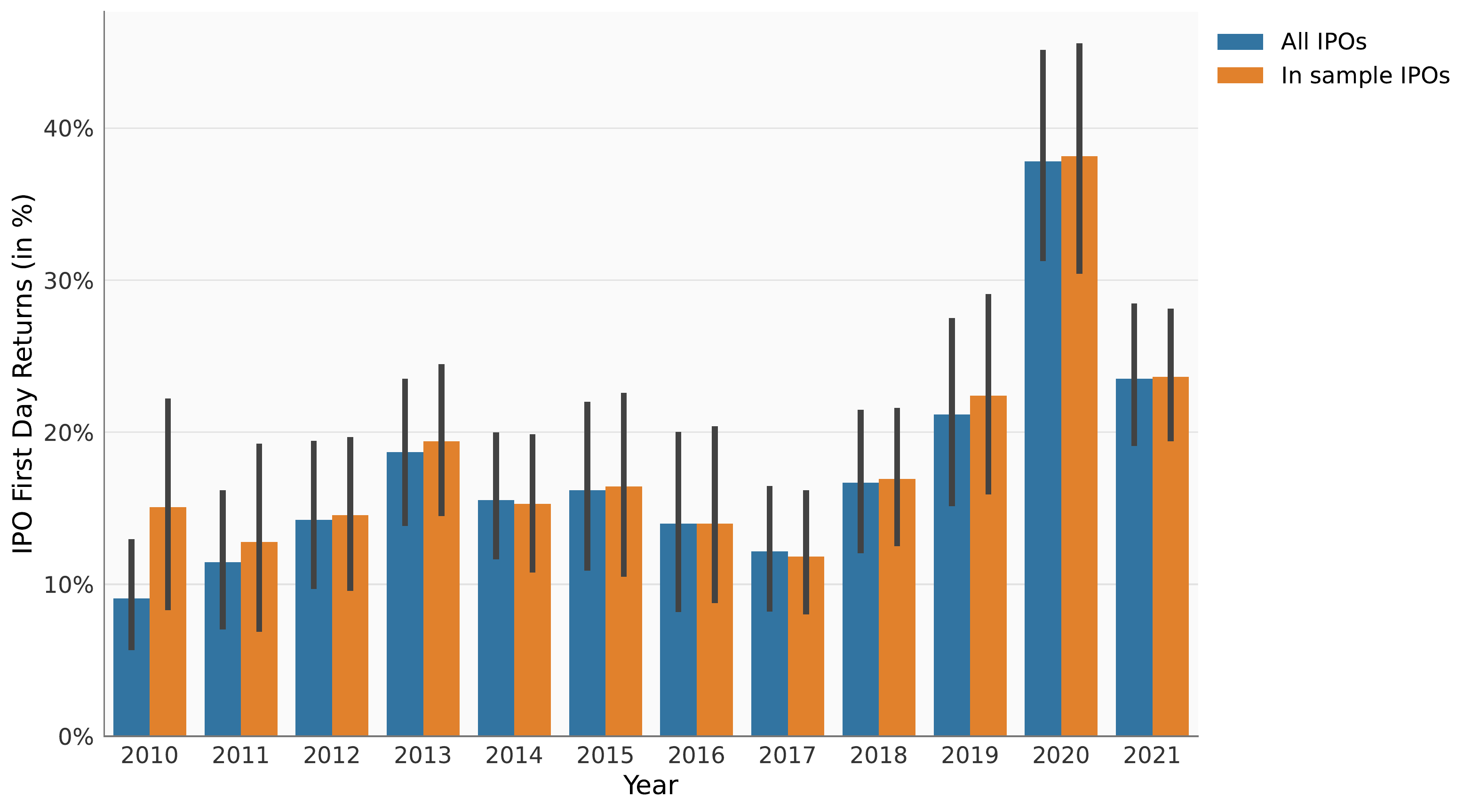}}

    \subfloat[]{\includegraphics[scale=0.33]{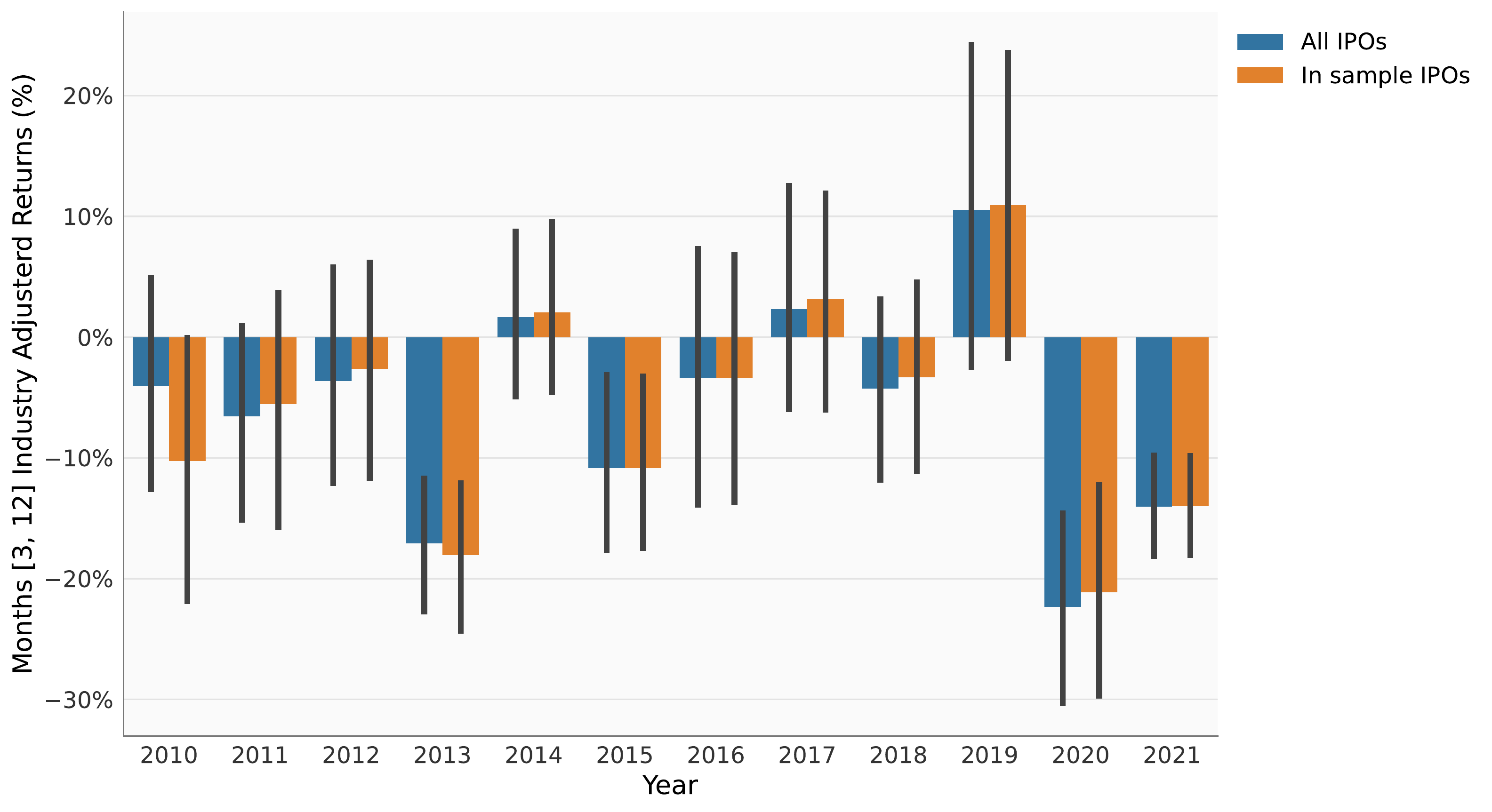}}

	\caption{Twitter's Coverage of IPOs}\label{fig:ipo_coverage_twitter}
\end{center}
 
\begin{flushleft}
	\footnotesize{Notes: Twitter provides extensive coverage of IPOs, making the sample representative of the broader IPO market.}
\end{flushleft}

\end{figure}

Figure \ref{fig:ipo_coverage_twitter} provides valuable insights into the coverage of IPOs on Twitter. Panel (a) displays a histogram that compares the number of IPOs in the matched sample to the total number of IPOs, indicating that the majority of IPOs are covered by social media. There was a slight decline in coverage during the early years of the study, which coincided with the initial stages of social media's popularity. However, from 2012 onwards, the coverage significantly increased, indicating that a large majority of IPOs are covered by social media. Panel (b) of the figure displays the average and standard deviation of first-day-returns for all IPOs and IPOs covered by social media, while Panel (c) plots the long-run industry adjusted returns. The plot shows that the averages and standard deviations of both sets of IPOs are similar and remain stable over time. This suggests that my sample is representative of the broader population of IPOs. The high coverage of IPOs in social media makes my findings more reliable and applicable to the general IPO market.

\subsubsection{Persistence of IPO Enthusiasm on Twitter}

I run regressions from Equation \ref{eq:autocorrelation} on the Twitter data. The results presented in Panel B of Table \ref{tab:autocorr_tw} are consistent with Table \ref{tab:autocorr}. Interestingly, I document a larger fraction of mostly bearish than mostly bullish users.

\begin{table}
    \centering
    \caption{Persistence of IPO Enthusiasm: Twitter \label{tab:autocorr_tw}}
    \begin{threeparttable}
    \begin{tabular}{lcccc} \hline 
        & \multicolumn{3}{c}{Number of IPOs covered} \\ \hline 
        & & & \\[\dimexpr-\normalbaselineskip+2pt] 
         & 3 & 4 & 5 \\ \hline 
         & & & \\[\dimexpr-\normalbaselineskip+2pt] 
        \textbf{Panel A: Valence Pattern} & & & \\
        & & & \\[\dimexpr-\normalbaselineskip+2pt] 
        Always Bullish & 374 & 127 & 58 \\
        Mostly Bullish & 1372 & 895 & 703 \\
        Always Bearish & 279 & 102 & 44 \\
        Mostly Bearish & 1547 & 1085 & 716 \\ \hline 
         & & & \\[\dimexpr-\normalbaselineskip+2pt] 
        \textbf{Panel B: Autocorrelation} & & & \\
        & & & \\[\dimexpr-\normalbaselineskip+2pt] 
        $\beta_2$ &  -0.1097 [0.0057] &  -0.0659 [0.0059] & -0.0372 [0.0060] \\
        $\beta_{2,i} > 0$ & 1214 & 751 & 552 \\
        $\beta_{2,i} < 0$ & 2347 & 1455 & 968 \\
        \hline \hline 
    \end{tabular}
        \begin{tablenotes}
            \small
            \item Note: This table delineates the persistence and evolution of investor valence across multiple IPOs. The ``Valence Pattern" panel identifies investors with consistent sentiment directions, categorizing them as either predominantly ``bullish" or ``bearish". The ``Autocorrelation" panel explores how an investor's sentiment toward an IPO might be informed by their valence from the previous IPO they engaged with. Specifically, I compute $\beta_3$ as specified in Equation \ref{eq:autocorrelation} with standard deviations in brackets. To get a sense of heterogeneity, I also compute $beta_{2,i}$, user-level autocorrelation and report the count of users with positive and negative coefficients. Source: Twitter.
        \end{tablenotes}
    \end{threeparttable}
\end{table}

\clearpage

\end{document}